\documentclass[11pt,a4paper]{article}

\usepackage{graphicx}
\usepackage{afterpage}
\usepackage{epsfig,cite}
\usepackage{amssymb}
\usepackage{amsmath}
\usepackage{dsfont}
\usepackage{multirow}
\usepackage{url,hyperref}
\usepackage{color}
\usepackage{ulem}

\usepackage{url}
\usepackage{hyperref}

\newcommand{\be}{\begin{equation}}
\newcommand{\ee}{\end{equation}}
\newcommand{\bea}{\begin{eqnarray}}
\newcommand{\eea}{\end{eqnarray}}
\newcommand{\bi}{\begin{itemize}}
\newcommand{\ei}{\end{itemize}}
\newcommand{\ben}{\begin{enumerate}}
\newcommand{\een}{\end{enumerate}}

\def\frac#1#2{{{#1}\over {#2}}}
\def\gsim{\mathrel{\rlap{\lower4pt\hbox{\hskip1pt$\sim$}}
    \raise1pt\hbox{$>$}}}         %greater than or approx. symbol
\def\lsim{\mathrel{\rlap{\lower4pt\hbox{\hskip1pt$\sim$}}
    \raise1pt\hbox{$<$}}}         %less than or approx. symbol

\newcommand{\draft}[1]{}

\def\beq{\begin{equation}}  
\def\eeq{\end{equation}}

\newcommand{\myMathSigma}[1]{\sigma_{\text{\tiny #1}}}
\newcommand{\mySigma}[1]{$\myMathSigma{#1}$}

\newcommand{\tabref}[1]{Table~\ref{#1}}
\newcommand{\tanb}[0]{\tan \beta}

\begin{document}
\begin{flushright}
Cavendish-HEP-14/09\\
PSI-PR-14-10\\
SLAC-PUB-16077\\
TTK-14-20
\end{flushright}
\begin{center}
{\Large \bf Improved cross-section predictions for heavy charged\\[2mm] Higgs boson production at the LHC}
\vspace{0.8cm}

{\bf 
Martin Flechl$^{1,2}$, Richard~Klees$^{3}$, Michael~Kr\"amer$^{3,4}$, Michael Spira$^{5}$,\\[1mm] and Maria~Ubiali$^{6,7}$
}

\vspace{1.cm}
{\it
~$^1$ Physikalisches Institut, Freiburg University,\\ 
Hermann-Herder Str. 3a, D-79104 Freiburg, Germany\\
~$^2$  Institute of High Energy Physics, Austrian Academy of Sciences,\\ 
Nikolsdorfergasse 18, A-1050 Vienna, Austria\\
~$^3$ Institute for Theoretical Particle Physics and Cosmology, RWTH Aachen University,\\ 
D-52056 Aachen, Germany\\
~$^4$ SLAC National Accelerator Laboratory, Stanford University, Stanford, CA 94025, USA\\
~$^5$ Paul Scherrer Institut, CH-5232 Villigen, Switzerland\\
~$^6$ Cavendish Laboratory, University of Cambridge,\\J.J. Thomson Avenue, CB3 0HE, Cambridge, UK\\
~$^7$ Department of Applied Mathematics and Theoretical Physics, \\
University of Cambridge, Wilberforce Road, CB3 0WA, Cambridge, UK}
\end{center}

\vspace{0.8cm}

\begin{center}
{\bf \large Abstract:}
\end{center}

In most extensions of the Standard Model, heavy charged Higgs bosons at the LHC are dominantly produced in association with heavy quarks. An up-to-date determination of the next-to-leading-order total cross section in a type-II two-Higgs-doublet model is presented, including a thorough estimate of the theoretical uncertainties due to missing higher-order corrections, parton distribution functions and physical input parameters. Predictions in the four- and five-flavour schemes are compared and reconciled through a recently proposed scale-setting prescription. A four- and five-flavour scheme matched prediction is provided for the interpretation of current and future experimental searches for heavy charged Higgs bosons at the LHC.

\clearpage

%%%%%%%%%%%%%%%%%%%%%%%%%%%%%%%%%%%%%%%%%%%%%%%%%%%%%%%%%%%%%%%

\section{Introduction}
\label{sec:intro}
Many extensions of the Standard Model (SM), in particular supersymmetric theories, 
require two Higgs doublets, leading to five physical scalar Higgs bosons, 
including two (mass-degenerate) charged particles $H^{\pm}$. 
Imposing natural flavour conservation, there are four different ways 
to couple the SM fermions to two Higgs doublets. Each of these four ways 
of assigning the couplings gives rise to a different phenomenology for 
the charged Higgs boson. Here the focus is on a type-II two-Higgs-doublet model 
(2HDM), in which one doublet generates the masses of the up-type quarks 
and the other of down-type quarks and charged leptons.

Searches at LEP have set a limit $m_{H^\pm} > 80$\,GeV on the mass of a charged Higgs 
boson for this model~\cite{hplep}.
For a branching ratio BR$(H^\pm \to \tau\nu)=1$, corresponding to the limit of large 
$\tan \beta$, the lower limit is 94\,GeV~\cite{hplep}. The Tevatron experiments place 
upper limits on BR$(t \rightarrow bH^{+})$ in the $(15-20)\%$ range for
$m_{\text{H}^+}<m_t$~\cite{hpcdf,hpd0} which have recently been superseded by 
results of the LHC experiments: 
preliminary ATLAS results~\cite{ATLAS-CONF-2013-090} for a type-II 2HDM exclude 
BR$(t \rightarrow bH^{+})$ larger than $(0.24-2.1)\%$ 
for 90\,GeV $< m_{H^\pm}<$ 160\,GeV and 
for the first time also provide cross section limits on $tH^\pm$ production in the mass 
range 180\,GeV $< m_{H^\pm}<$ 600\,GeV, 
both with the assumption that BR$(H^\pm \to \tau\nu)=1$. Based on the same assumptions, 
CMS results~\cite{hpcms} exclude BR$(t \rightarrow bH^{+})$ above $(2-4)\%$ for 
charged Higgs boson masses between 80 and 160\,GeV. 
The search for a charged Higgs boson is a central part of the physics program 
at the LHC, and a discovery would provide 
unambiguous evidence for an extended Higgs sector beyond the SM. 

In this paper the focus is on heavy charged Higgs bosons, $m_{H^{\pm}} > m_t$. 
Their main production mechanism at the LHC in most extensions of the SM 
proceeds through associated production 
with a top quark, 
\begin{equation} 
pp\, \rightarrow\, tH^{\pm}(b)+X.
\end{equation}

In a two-Higgs doublet model of type II,  like the Minimal Supersymmetric extension of the SM (MSSM), 
the Yukawa coupling of the charged Higgs boson $H^-$ to a top quark and bottom antiquark is given by 
\begin{equation}
g_{t\bar{b}H^-} = \sqrt{2}\left(\frac{m_t}{v}P_R\cot\beta + \frac{m_b}{v}P_L\tan\beta\right),
\label{eq:tbhcoupling}
\end{equation}
where $v = \sqrt{v_1^2+v_2^2} = (\sqrt{2}G_F)^{-\frac12}$ is the Higgs vacuum expectation value in the 
SM, with the Fermi constant $G_F = 1.16637 \times 10^{-5}\;\mbox{GeV}^{-2}$~\cite{PDG}. 
The parameter $\tan\beta = v_2/v_1$ is the ratio of the vacuum expectation values $v_1$ and $v_2$ 
of the two Higgs doublets, and $P_{R/L} = (1\pm\gamma_5)/2$ are the chirality projectors. 

There exist two ways of calculating the associated production of charged Higgs bosons 
with a top quark and an untagged bottom quark. One option, which is straightforward from 
the conceptual point of view, is to consider the bottom-quark mass to be of the same 
order of magnitude as the other hard scales involved in the process. Then, the bottom 
quark does not contribute to the proton wave function and can only be generated as a 
massive final state. In practice, the theory which is used in such a calculation is 
an effective theory with four light quarks, where the bottom quarks are decoupled and 
do not enter the computation of the running coupling constant and the evolution of 
the parton distribution functions (PDFs). According to this approach, named four-flavour scheme (4FS), 
the lowest-order QCD production processes are gluon-gluon fusion and quark-antiquark annihilation, 
$gg \rightarrow tbH^\pm$ and $q\bar{q} \rightarrow tbH^\pm$, respectively. 
The former is dominant at the LHC due to the large gluon-gluon luminosity.
In the four-flavor scheme, computations are more involved due to the 
higher final-state multiplicity and because the 
additional final-state particle is massive. 
However, the kinematics of the heavy 
quark is correctly taken into account already at the leading order, and the interface 
with parton shower codes is straightforward. The drawback is that potentially 
large logarithms of the ratio of the hard scale of the process and the mass of the 
bottom quark, which arise from the splitting of incoming gluons into nearly collinear 
$b\bar{b}$ pairs, are not summed to all orders in perturbation theory.

Such a summation is achieved in the so-called five-flavour scheme (5FS), in which 
the bottom-quark mass is considered to be much smaller than the other scales 
involved in the process and consequently is ignored. The bottom quarks are 
treated as massless partons which are constituents of the proton and may thus appear 
in the initial state.
Using bottom-quark PDFs requires the approximation that, 
at leading order, the outgoing $b$ quark has small transverse momentum 
and is massless, and that the virtual $b$ quark is quasi on shell. 
In this scheme, the logarithms associated to the initial-state collinear splitting
are resummed to all orders in perturbation theory by means of the Dokshitzer-Gribov-Lipatov-Altarelli-Parisi  
evolution of the bottom parton densities \cite{Barnett:1987jw,Dicus:1988cx}. 
The leading-order (LO) process for the inclusive $tH^\pm$ cross section is 
gluon-bottom fusion, $gb \rightarrow tH^\pm$. The next-to-leading-order (NLO) 
cross section in the 5FS comprises $\mathcal{O}(\alpha_s)$ corrections 
to $gb \rightarrow tH^\pm$, including the LO contributions of the 4FS calculation,
 $gg \rightarrow tbH^\pm$ and $\bar{q}q \rightarrow tbH^\pm$.
The $m_b=0$ approximation in the 5FS can be systematically improved by introducing 
$m_b\neq 0$ in higher-order contributions corresponding to diagrams where the $b$ 
quark only appears in the final state (see for example Ref.~\cite{Tung:2001mv} and references therein). 

To all orders in perturbation theory the improved 5FS and the 4FS are identical, 
but the way of ordering the perturbative expansion is different, and at a finite order, 
results do not match exactly. For some processes the difference between calculations performed 
in the two schemes was found to be very significant at leading order. 
One of the most striking examples is the discrepancy which was initially observed 
in inclusive neutral Higgs boson production initiated by $b$ quarks; see Ref.~\cite{Campbell:2004pu}. 
In the leading-order analysis, setting the renormalization and factorization scales to $\mu_f = m_H$, 
the 5FS prediction exceeds the 4FS prediction by more than a factor of 5. 
This has led to several thorough studies aiming to shed light on the origin of 
this difference~\cite{Dicus:1988cx,rainwater-spira,MKRAEMER,tilman:higgs,maltoni:higgs}. 
It has been shown that a choice 
such as $\mu_f=m_H/4$ leads to a reduced scale 
dependence in both approaches and that the discrepancy between the
schemes was reduced~\cite{Kramer:2010zz,Dittmaier:2010zz}, 
thus suggesting that the scale at which the gluon splits is softer than the scale of 
the hard process where the Higgs boson is produced. With this scale choice, the four- and five-flavour scheme 
calculations numerically agree within their respective uncertainties once higher-order QCD corrections are 
taken into account. 

A recent analysis presented in Ref.~\cite{MRU2012} investigates the dynamical origin of 
such a scale choice for generic processes involving bottom quarks in the initial state at the LHC.
It is shown, contrary to na\"ive expectations, that unless the mass of the produced particles 
is very large, the effect of initial-state collinear logarithms involving the effective scale 
${\cal Q}$ and the bottom-quark mass, $\log({\cal Q}^2/m_b^2)$, is always
modest. 
Even though total cross sections computed in the five-flavor scheme exhibit a smaller scale uncertainty, 
such initial state collinear logarithms do not spoil the convergence of perturbation theory in four-flavor 
scheme calculations. One of the reasons of this perturbative behaviour is that the effective scale 
${\cal Q}$ which enters the initial-state collinear logarithms is significantly smaller than the 
hardest scale in the process. The effective scale is modified by universal phase-space 
factors that tend to reduce the size of the logarithms for processes taking place at hadron colliders. 
This provides a simple rule to choose the factorization scale at which to perform comparisons 
between calculations in the four- and five-flavor schemes. The scale turns out to be similar to the previous 
choice of $\mu_f=m_{H^\pm}/4$ for the bottom quark PDF in the 5FS, based on considerations
on the transverse momentum of the bottom quark~\cite{PL2002}.

If the charged Higgs boson is discovered at a high mass,  $m_{H^{\pm}} \gg m_t$, then it will be instructive 
to assess the impact 
of the resummation of the collinear logs $(m_{H^\pm}/m_t)$ into top-quark PDFs and comparing 
the five- to the six-flavor scheme in one of the possible scenarios of a future 100\,TeV collider; see e.g.\
Ref.~\cite{Dawson:2014pea}. 

\medskip

Next-to-leading order predictions for heavy charged Higgs boson
production at the LHC in a type-II two-Higgs-doublet model 
have been made in the past in both 5F~\cite{Zhu:2001nt,Gao:2002is,PL2002,Berger:2003sm,Kidonakis:2004ib,Kidonakis:2005hc,Weydert:2009vr,Kidonakis:2010ux} 
and 4F schemes~\cite{Peng:2006wv,MKRAEMER}, 
including also electroweak corrections~\cite{Beccaria:2009my,Nhung:2012er}. 
In Refs.~\cite{Kidonakis:2004ib,Kidonakis:2005hc,Kidonakis:2010ux}
the approximate next-to-next-to-leading-order 5FS cross section including the 
next-to-leading-log (NLL) and next-to-next-to-leading-log (NNLL) soft gluon logarithmic corrections 
are presented in
the charged Higgs boson mass range from 200 GeV to 1000 GeV. Soft gluon resummation 
enhances the NLO 5FS cross section and stabilizes it under scale variation.
It is also stated that NNLL corrections further enhance the cross section compared to the NLO
result, up to 15\% in the mass range that is considered here and with the central scale choices 
made by the author. For softer central scale choices, it is expected that soft gluon corrections 
enhance the NLO cross section by a smaller factor such that the enhanced resummed cross section 
still lies within the scale uncertainty (which is about 10\%). 
Resummed calculations are not available in the four-flavor scheme.

In the work presented here, the NLO-QCD predictions are updated and improved by adopting the new scale setting 
procedure~\cite{MRU2012} mentioned above.  
A thorough account of all sources of theoretical uncertainties is given, state-of-the-art PDF sets
are used, and uncertainties are consistently combined.
A matched prediction~\cite{SANTANDER} for the four- and five-flavour scheme calculations is provided. 
Furthermore, results for a large range of $\tan\beta$ are presented which allow the comparison 
between theory and experiment for a wide class of beyond-the-SM scenarios. This work has been performed within the 
LHC Higgs Cross Section Working Group (LHC-HXSWG)~\cite{HXSWG}, and first preliminary results have been presented 
in Ref.~\cite{yr3}. 

%%%%%%%%%%%%%%%%%%%%%%%%%%%%%%%%%%%%%%%%%%%%%%%%%%%%%%%%%%%%%%%%%%
\section{Theoretical Settings}
\label{sec:settings}
In Secs. \ref{sec:5F} and \ref{sec:4F}, the total cross sections for associated top quark 
and charged Higgs boson production are calculated using the five- and the four-flavour schemes, respectively. 
In this section the generic settings of the two calculations are specified, and the
method of estimating the theoretical uncertainty is presented. 

All cross sections are computed for a type-II 2HDM, for which the coupling between a charged Higgs 
boson, a bottom antiquark and a top quark is given in Equation~\eqref{eq:tbhcoupling}.
\footnote{Charged Higgs boson production in a type-I 2HDM is discussed in Sec.~\ref{sec:tanbeta}.}
The only parameters that enter the calculation are thus 
the particle masses and $\tan\beta$, so that the results are rather generic. 
However, for supersymmetric models, additional higher-order contributions 
through the virtual exchange of supersymmetric particles need to be included. 
Corrections that modify the tree-level relation between 
the bottom quark mass and its Yukawa coupling are of particular relevance. 
These corrections are enhanced at large $\tan\beta$ 
and can be summed to all orders through a modification of the Yukawa coupling; see
Refs.~\cite{MKRAEMER,Carena:1993bs,Pierce:1996zz,Hempfling:1993kv,
Hall:1993gn,Guasch:2003cv,Carena:1999py,Noth:2010jy,Noth:2008tw,Mihaila:2010mp}. 
The remaining supersymmetric (SUSY)-QCD effects are marginal at large $\tan\beta$, 
but can reach up to {\cal O(10\%)} at small $\tan\beta$. 
Specifically, for the benchmark point SPS1b~\cite{Allanach:2002nj} 
the SUSY-QCD effects beyond those contained in the Yukawa coupling amount 
to $-(6/8/5/0.1)$\% for $\tan\beta = 3/5/10/30$, respectively. 
Furthermore, since resummed results including soft gluon resummation 
are available at NLO+NLL and NLO+NNLL accuracy only for the 5FS cross 
section~\cite{Kidonakis:2004ib,Kidonakis:2010ux} and not in the 4FS, 
for consistency they are not included. These contributions enhance the 5FS cross section 
independently of the value of $\tan\beta$. Since they are expected to have a similar effect on the 4FS 
cross section, their exclusion does not invalidate the results of the comparison presented in this analysis. 
These corrections beyond NLO are within the estimated scale uncertainties and are thus accounted for 
by this more conservative estimate of theoretical uncertainties.

Note that throughout this paper results are given for the $tH^-$ final state. 
The charge conjugate final state $\bar{t}H^+$ can be included by multiplying the results presented here 
by a factor of 2. 

\medskip

There are two different sources of theoretical uncertainties that need to be taken 
into account: scale uncertainties which should reflect the error due to the 
omission of corrections beyond NLO, and parametric uncertainties induced by the 
error on the input parameters. 

The scales that enter the calculation of heavy charged Higgs boson production 
comprise the renormalization scale $\mu_r$ which determines the running of $\alpha_s$, 
the factorization scale $\mu_f$ which determines the evolution of the parton distribution functions, 
and the scale $\mu_b$ which determines the running bottom quark mass 
in the Yukawa coupling. 
To estimate the overall scale uncertainty, all three scales are independently varied by a 
factor of 2 about their central value, as specified 
in Secs.~\ref{sec:5F} and \ref{sec:4F} for the five- and four-flavor schemes, respectively. 
To avoid spurious large logarithms, scale choices 
where the ratio of any of the three scales exceeds a factor of 2 are not taken into account. 
The envelope of the predictions is used
to determine the overall scale uncertainty. 

The input parameters relevant for heavy charged Higgs boson production are the parton distributions functions, 
the strong coupling 
$\alpha_s$ and the bottom-quark mass $m_b$. 
The impact of the top-quark mass uncertainty on the results is negligible and thus not considered. 
These uncertainties are correlated as the PDF fits are performed for specific values of $\alpha_s$ 
and $m_b$, so care has to be taken to estimate 
the input parameter uncertainties in a consistent way. 
The computer codes employed to calculate the next-to-leading order total cross section in the two 
schemes have been interfaced to the LHAPDF library~\cite{LHAPDF}. This made it possible 
to use modern PDF sets and consistent $\alpha_s(M_Z)$ values
in the PDFs and in the matrix element computation.
In this analysis the charged Higgs boson cross sections are determined by using three 
of the most recent 
PDF sets, determined from global analyses of Deep-Inelastic Scattering (DIS) and hadron collider data, namely 
CT10~\cite{CT10}, MSTW2008~\cite{MSTW2008} and NNPDF2.3~\cite{NNPDF23}. 
The latter set already includes the constraints coming from the early LHC data. 
The default General-Mass Variable 
Flavour Number (GM-VFN) sets are used for the computation of the 5FS cross section\footnote{Both 
CT10 and MSTW2008 evolve $\alpha_s$ with $n_f=5$ at all scales, while NNPDF2.3 evolves $\alpha_s$ 
with $n_f=6$ above the top threshold. 
For this computation it has been checked that freezing $n_f=5$ above the top threshold 
in the predictions obtained with NNPDF2.3 does not affect the result.}, 
while the Fixed Flavour Number (FFN)
sets with $n_f=4$ are used in the computation of the 4FS cross sections. 
In the GM-VFN scheme, a parton distribution function is associated to all partons, including the bottom quarks, 
above production 
threshold. The mass of the heavy quarks is
taken into account in the DIS partonic cross sections. The GM-VFN scheme is 
designed to interpolate between the FFN scheme, which gives a correct
description of the threshold region, and the resummation of the large collinear logarithms at large $Q^2$.
Each PDF set adopts a variant of the GM-VFN scheme, which differs by higher-order terms associated to the 
matching condition. In particular, CT10 adopts a variant of the ACOT scheme~\cite{ACOT} called
ACOT-$\chi$~\cite{Tung:2001mv}. The Thorne-Roberts~\cite{TR} 
VFN scheme or TR'~\cite{Thorne:2006qt} in its latest version, which emphasizes
the correct threshold behaviour and includes certain higher-order terms, 
is adopted in the MSTW2008 PDF determination. 
The fixed-order-next-to-leading-log (FONLL) approach, introduced in Ref.~\cite{FONLL} in the context of
hadro-production of heavy quarks and more recently applied to deep-inelastic
structure functions~\cite{FONLLdis},
considers both massless and massive scheme calculations as power expansions
in the strong coupling constant, and replaces the coefficient of the
expansion in the former with their exact massive counterparts. The FONLL approach is adopted by the
NNPDF collaboration.

In addition to the default GM-VFNS PDF sets, each collaboration provides a FFN 
scheme set with $n_f=4$~\cite{MSTW2008nf4,NNPDF23FFN}, which allows for a theoretically 
consistent cross section prediction in a 4FS calculation. Other PDF fitting collaborations~\cite{Alekhin:2013nda}
provide as default FFNS parton sets with $n_f=3,4,5$. 

When the corresponding PDF set is available, a common value of $\alpha_s$ is chosen for the predictions
according to the recent PDG average~\cite{PDG}: $\alpha_s(M_Z)=0.1183\pm0.0012$. 
The PDF+$\alpha_s$ uncertainty is estimated by 
using the sets determined at different values of $\alpha_s(M_Z)$
provided by each collaboration, and by combining the PDF and the $\alpha_s$ uncertainties according
to the standard prescription, as illustrated for example in Sec. 3.2 of Ref.~\cite{Alekhin:2011sk}.

The pole mass of the top and bottom quarks are set to $m_t$ = 172.5\,GeV and 
$m_b$ = 4.75\,GeV, respectively.
The choice of the bottom quark mass and the corresponding uncertainty deserve careful consideration.
The calculation of hadronic cross sections always involves PDFs which have
an intrinsic dependence on the mass of the bottom quark.
The central value for the bottom pole mass adopted here is consistent with most PDF fits, 
and corresponds to a 
$\overline{\rm MS}$ mass of $m_b(m_b)=4.21$\,GeV, using the two-loop QCD 
relation~\cite{Kuhn:2007vp}\footnote{Note that for consistency we use the one-loop QCD relation to convert 
the pole mass into the running bottom mass, obtaining $m_b(m_b)=4.3378$}. 
This value is close but not identical  
to the current PDG value, $m_b(m_b)=4.18$\,GeV~\cite{PDG}, and to the
recommendation from the LHC-HXSWG, $m_b(m_b)=4.16$\,GeV. %} 
The uncertainty due to $m_b$, in particular the dependence of the PDFs on the bottom-quark mass,  
is investigated by using input sets of PDFs with $m_b$ varied by 60 MeV about its central 
pole mass value. As shown in Ref.~\cite{MSTW2008nf4} and in the recent study of bottom quark-initiated
neutral Higgs boson production~\cite{Bagnaschi:2014zla}, the 
bottom-quark PDF exhibits a strong dependence on the bottom-quark mass adopted in the PDF fit. 
Thus a significant dependence of the 5FS predictions on $m_b$ through the bottom-quark PDF is expected. 
In addition, the cross section for charged Higgs boson production depends on
the bottom quark mass through the bottom quark Yukawa coupling and, 
for the 4FS calculation, through the explicit $m_b$-dependence 
in the matrix element. All these uncertainties have been included in this calculation. 
Note that the bottom quark mass dependence of the 4F PDFs is small as $m_b$ enters 
only indirectly through the DIS cross section in the global fit. 
However, the matrix element in the 4F scheme calculation contains a $m_b$-dependent collinear 
logarithm, which corresponds to the bottom-quark PDF in the 5FS.

In summary, the total theoretical uncertainty quoted here includes: 
the PDF uncertainty $\delta_{\rm PDF}$, 
the $\alpha_s$ and $m_b$ uncertainties ($\delta\alpha_s$ and $\delta m_b$), 
and the uncertainty due to missing higher orders in the partonic cross section, $\Delta^{\pm}_{\mu}$, 
estimated according to the usual procedure by varying the three different scales $\mu_r$, $\mu_f$ and $\mu_b$ 
by a factor of 2 about their central values, as described above. 

To combine the various sources of theoretical uncertainty the prescription of 
the LHC-HXSWG is used. The combined PDF$+\alpha_s+m_b$ uncertainty for each different PDF input 
set is computed first\footnote{Unfortunately, not all PDF sets allow one to vary $\alpha_s$ or $m_b$, 
as specified in 
the following sections. Therefore the total PDF uncertainty may be slightly underestimated.}. 
For each PDF set {\it i} the three sources of uncertainty are 
combined in quadrature~\cite{Alekhin:2011sk}:
\begin{equation}\label{eq:uncertainty}
\delta_{{\rm PDF}+\alpha_s+m_b}^{i} = \sqrt{(\delta^i_{\rm PDF})^2 + (\delta^i\alpha_s)^2 + (\delta^i m_b)^2}.
\end{equation}
At this point, the envelope of the three predictions is used to give an estimate of the combined
PDF and parametrical uncertainty $\delta_{{\rm PDF}+\alpha_s+m_b}$. Following 
Ref.~\cite{Alekhin:2011sk} the central prediction is 
defined as the midpoint of the envelope, such that the PDF$+\alpha_s+m_b$ uncertainty is 
symmetric by construction. 
The scale uncertainty, estimated for the central choice of input parameters, 
is then added linearly to the  combined PDF$+\alpha_s+m_b$ uncertainty~\cite{Dittmaier:2011ti}:
\begin{equation}\label{eq:tot-uncertainty}
\Delta_{\rm tot}^{\pm} = \pm\delta_{{\rm PDF}+\alpha_s+m_b} 
\pm \Delta_{\mu}^{\pm}\,.
\end{equation}
Note that the scale uncertainty does not significantly depend on the choice of PDFs, and is  
computed by using the central {\tt CT10} set. 

Cross sections are calculated for Higgs boson masses $m_{H^\pm}$ in the range from 200\,GeV to 600\,GeV 
in steps of 20\,GeV. Detailed results for $m_{H^\pm}=200,400$ and 600\,GeV are collected 
in Tables~\ref{tab:5F:pdf}--\ref{tab:4F:pdf}.  
The value of $\tanb$ is set to 30, in correspondence to the region favoured by 
recent MSSM fits~\cite{MSSM_EXCLUSION}. 
In Sec.~\ref{sec:tanbeta} the cross section as a function of $\tanb$ is presented. 

%%%%%%%%%%%%%%%%%%%%%%%%%%%%%%%%%%%%%%%%%%%%%%%%%%%%%%%%%%%%%%%%%%

\section{Five-flavour scheme results}
\label{sec:5F}

In the five-flavor scheme, bottom quarks are treated as massless partons which 
appear in the initial state. The leading-order (LO) process for the inclusive $tH^\pm$ 
cross section is gluon-bottom fusion, $gb \rightarrow tH^\pm$. 
The next-to-leading order (NLO) QCD cross section in the 5FS has been calculated in 
Refs.~\cite{Zhu:2001nt,Gao:2002is,PL2002,Berger:2003sm,Kidonakis:2005hc,Weydert:2009vr}.
All numbers presented here have been computed by interfacing the public code 
Prospino~\cite{Beenakker:1996ed, Prospino2} with the LHAPDF library~\cite{LHAPDF}.

The renormalization scale is set to the average final state mass
$\mu_r$ = $(m_{H^\pm}+m_t)/2$. As previously discussed, 
the factorization scale is set according to the method proposed in Ref.~\cite{MRU2012}. 
There, a simple analytic formula is provided, which enables a quantitative assessment 
of the size of the collinear logarithms 
resummed in a 5FS computation. Hence, a factorization scale can be chosen to optimally 
perform comparisons between calculations in the four- and five-flavor schemes. 
For processes at the LHC this scale is typically smaller than the hard scale, 
since the effective scale entering the 
initial-state collinear logarithms is damped by a kinematic factor which depends on the 
final-state phase space. For charged Higgs boson 
production, the scale associated to the gluon splitting into bottom quarks is
\begin{equation}
\begin{array}{lclcl}
\mathcal{Q}_{tHb}^2 = M^2 \frac{\displaystyle \left(1-z\right)^2}{\displaystyle z} & \, & \text{with} & \, & z = \frac{\displaystyle M^2}{\displaystyle \hat{s}},
\end{array}
\end{equation}
where $M^2=(m_{H^\pm} + m_t)^2$ and $\hat{s}$ is the partonic center-of-mass energy. 
By weighting this event-by-event logarithmic factor with the hard matrix element and the 
luminosity, a constant scale $\tilde{\mu}_f$ can be estimated which only depends on 
$m_{H^\pm}$, $m_t$ and on the collider center-of-mass energy $\sqrt{s}$. 
At this scale, the 5FS prediction can be meaningfully compared to the one
in the 4FS~\cite{MRU2012}. 
The factorization scale $\tilde{\mu}_f$ is presented in Table~\tabref{fac_scale:result} 
for the full range of Higgs boson masses considered, for center-of-mass energies of
$\sqrt{s}=8$ and 14\,TeV. 

%%%%
\begin{table}
\renewcommand*{\arraystretch}{1.2}
\begin{center}
\begin{tabular}{ |r|r|r|r|r| }
  \hline
  &\multicolumn{2}{|c|}{8\,TeV} & \multicolumn{2}{|c|}{14\,TeV}\\
  \hline
$m_{H^\pm}$ (GeV) & $\tilde{\mu}_f$ (GeV) & $(m_{H^\pm}+m_t)/\tilde{\mu}_f$ & $\tilde{\mu}_f$ (GeV)& $(m_{H^\pm}+m_t)/\tilde{\mu}_f$ \\
\hline
 200 & 68.9  & 5.5 & 76.3  & 4.9 \\
 220 & 70.7  & 5.6 & 79.6  & 4.9 \\
 240 & 73.4  & 5.6 & 82.7  & 5.0 \\
 260 & 75.9  & 5.7 & 85.9  & 5.0 \\
 280 & 78.5  & 5.8 & 89.0  & 5.1 \\
 300 & 81.0  & 5.8 & 92.0  & 5.1 \\
 320 & 83.5  & 5.9 & 95.0  & 5.2 \\
 340 & 85.9  & 6.0 & 97.9  & 5.2 \\
 360 & 88.3  & 6.0 & 100.9 & 5.3 \\
 380 & 90.7  & 6.1 & 103.8 & 5.3 \\
 400 & 93.0  & 6.2 & 106.6 & 5.4 \\
 420 & 95.3  & 6.2 & 109.4 & 5.4 \\
 440 & 97.5  & 6.3 & 112.2 & 5.5 \\
 460 & 99.7  & 6.3 & 115.0 & 5.5 \\
 480 & 101.9 & 6.4 & 117.7 & 5.5 \\
 500 & 104.0 & 6.5 & 120.4 & 5.6 \\
 520 & 106.2 & 6.5 & 123.1 & 5.6 \\
 540 & 108.2 & 6.6 & 125.7 & 5.7 \\
 560 & 110.3 & 6.6 & 128.3 & 5.7 \\
 580 & 112.3 & 6.7 & 130.9 & 5.7 \\
 600 & 114.3 & 6.7 & 133.4 & 5.8 \\
  \hline
\end{tabular}
\end{center}
\caption{Dynamical factorization scale $\tilde{\mu}_f$ for $pp\rightarrow  t H^\pm + X $ for 
the LHC at $\sqrt s =8$ and 14\,TeV. The four-flavour {\tt CT10nlo\_4f} PDF set has been used as input 
to evaluate $\tilde{\mu}_f$ according to Eq.~(5.13) of Ref.~\cite{MRU2012}.
\label{fac_scale:result}
}
\end{table}

\begin{figure}
\begin{center}
        \includegraphics[width=0.48\textwidth]{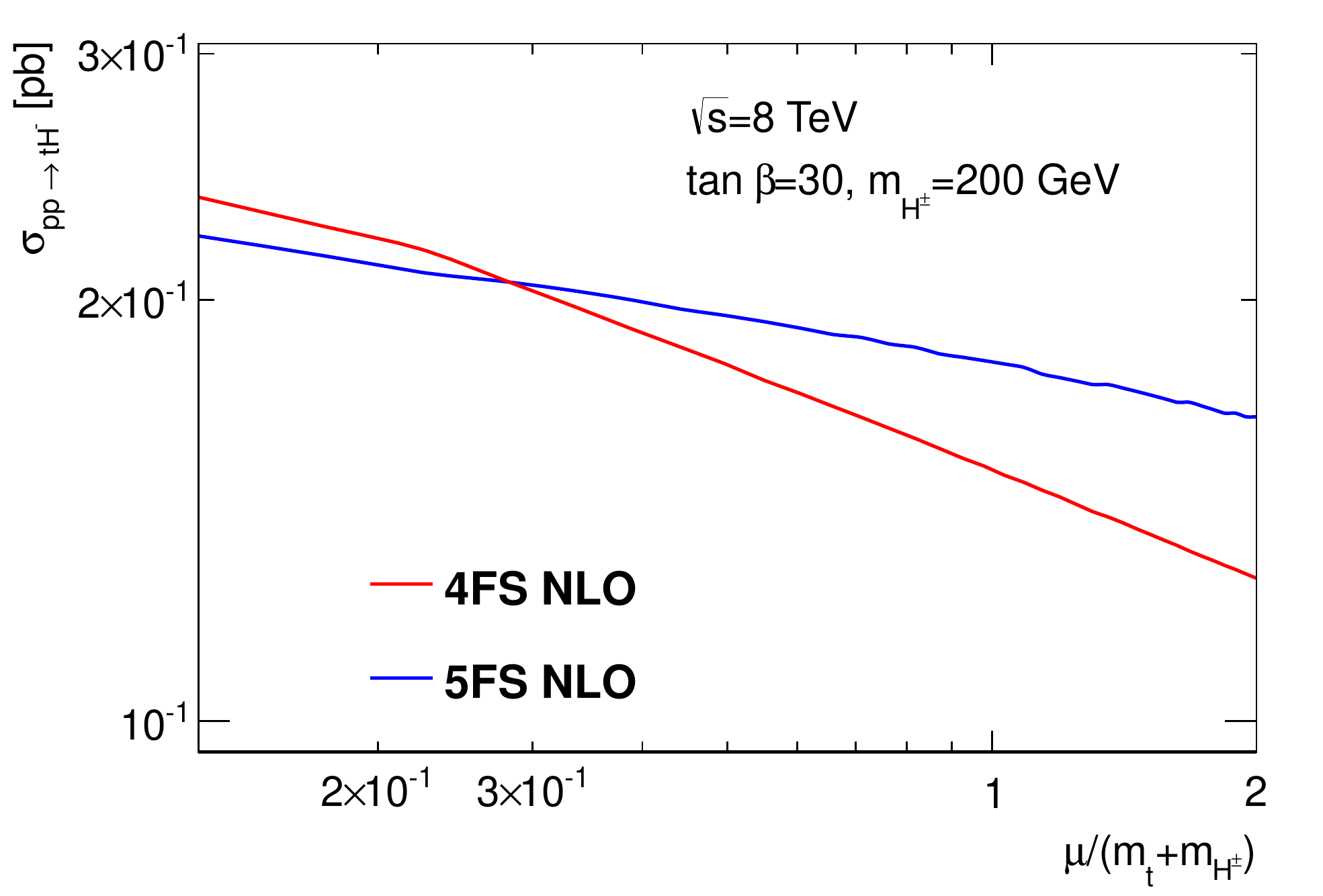}
        \includegraphics[width=0.48\textwidth]{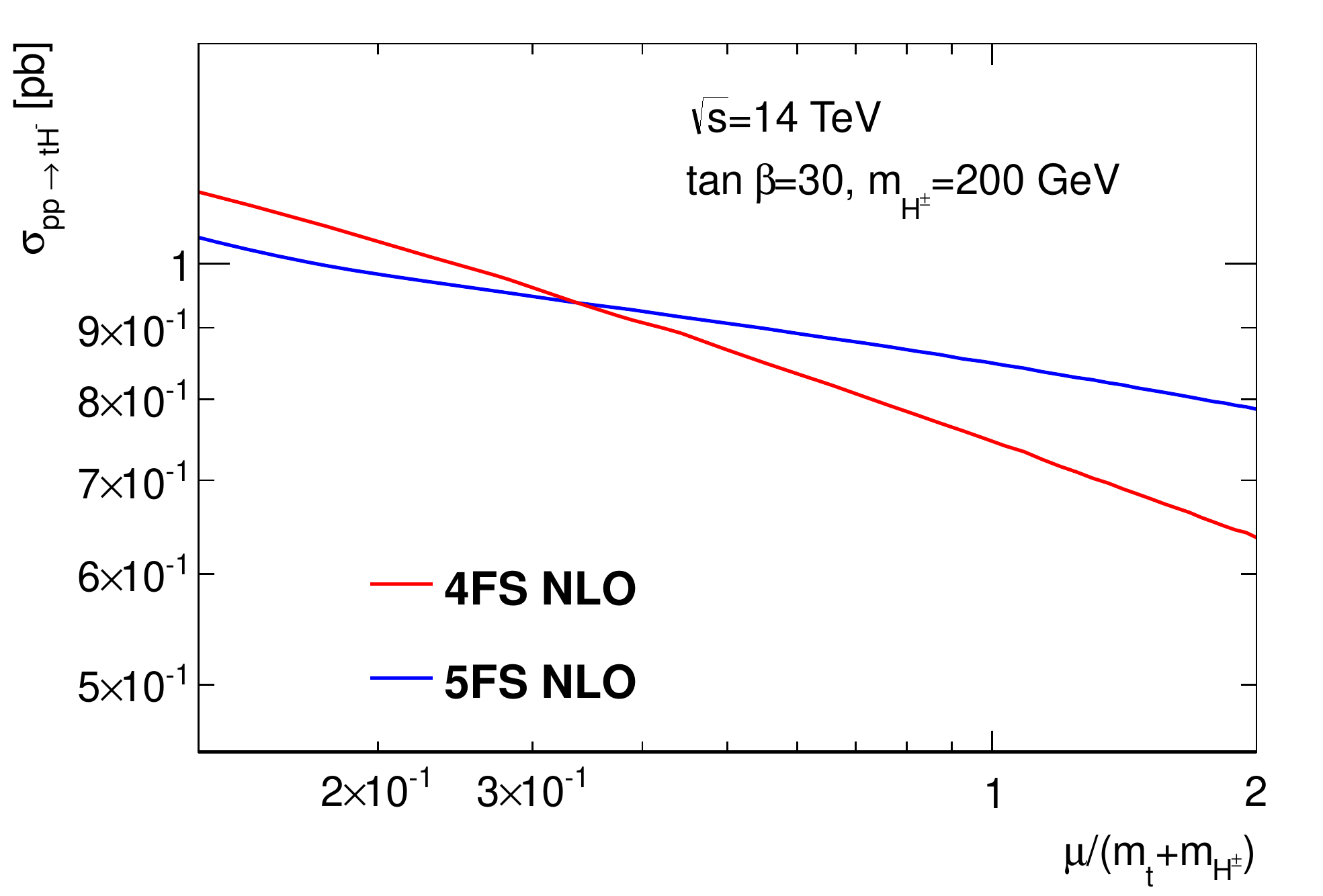}
        \includegraphics[width=0.48\textwidth]{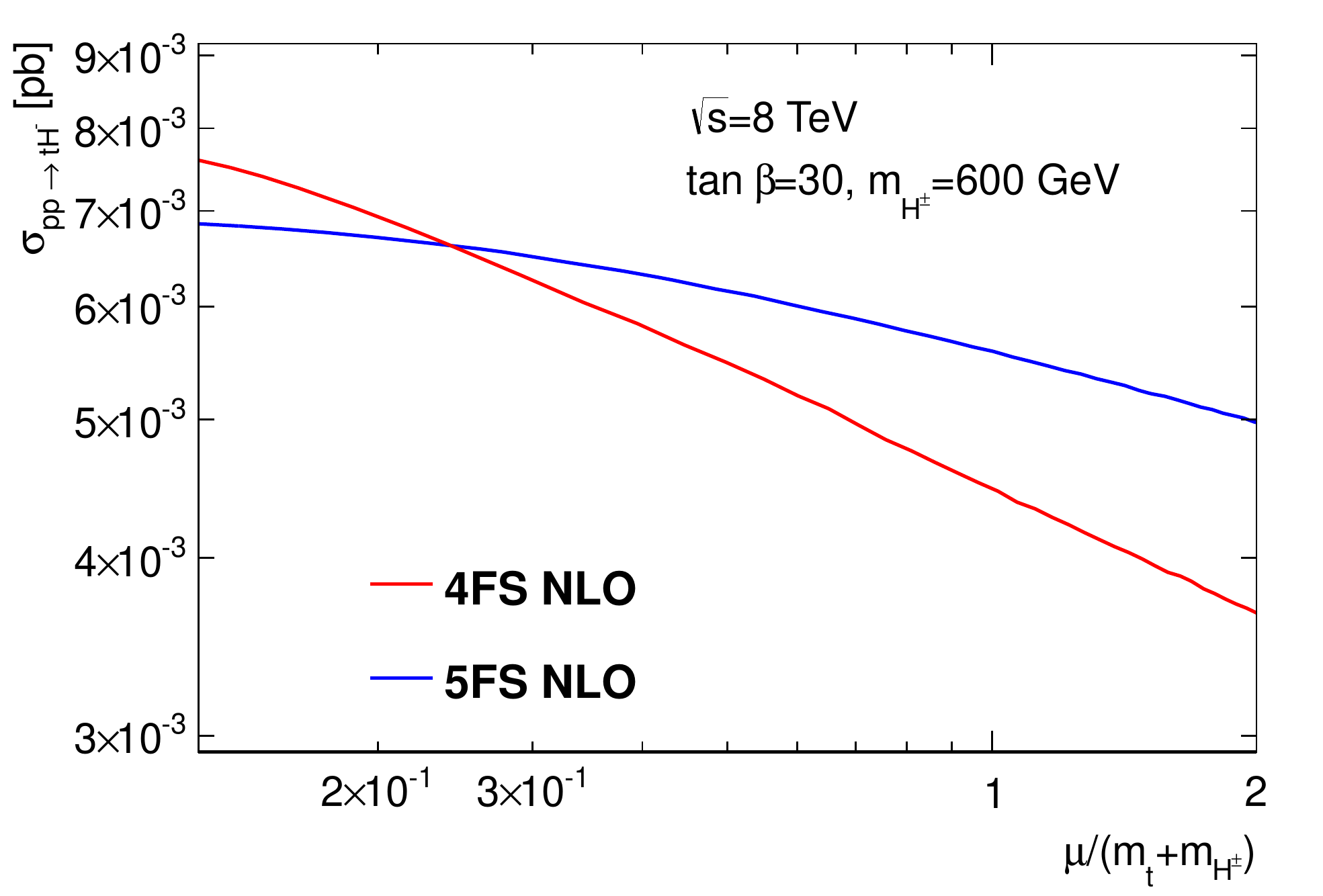}
        \includegraphics[width=0.48\textwidth]{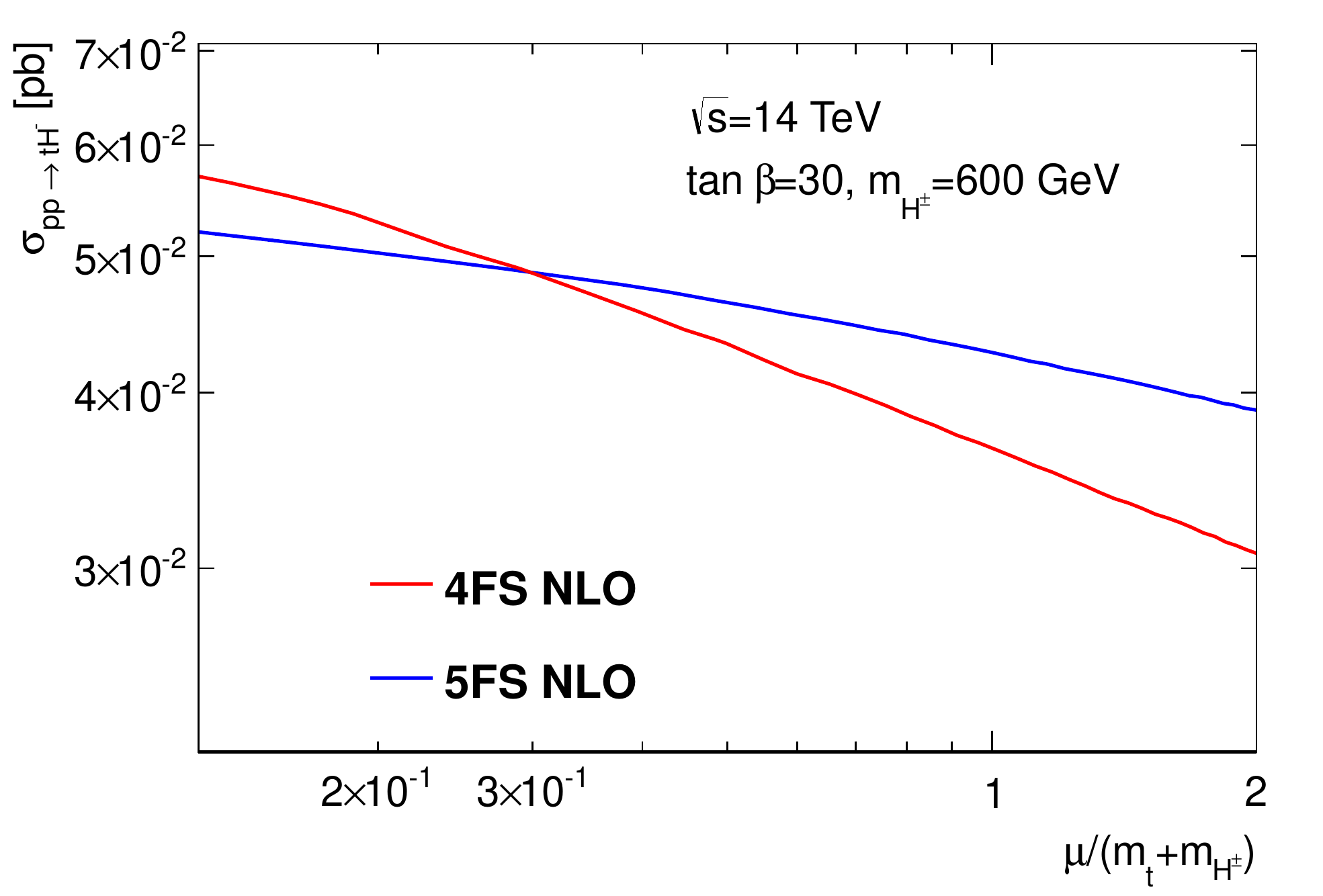}
\end{center}
        \caption{The dependence of the four- and five-flavor scheme $pp \to tH^- +X$ total cross sections on the 
factorization and renormalization scales, for $m_H^\pm=200$ (top) and $600$\,GeV (bottom) 
and $\sqrt{s}=8$\,TeV (left) and $14$\,TeV (right). 
Both scales are varied simultaneously between ${\mu}/10$ and $2\mu$ about $(m_{H^\pm}+m_t)$.}
        \label{fig:scale}
\end{figure}

The dependence of the total cross section on the renormalization and factorization scales 
is illustrated in Fig.~\ref{fig:scale}, for the largest and smallest $m_{H^\pm}$ values  
considered in this analysis. For the sake of illustration, the same value is used for both scales. 
This comparison, analogously to the one shown in Refs.~\cite{MRU2012,maltoni:stop1},
is meant to illustrate the overall dependence of the total cross section on the scales 
that enter the computation. It is not meant to provide an exact estimate of the scale uncertainty.
Both renormalization and factorization scales are varied between ${\mu}/10$ and $2\mu$ around $(m_{H^\pm}+m_t)$,
which is the natural hard scale of the process.
For comparison, in Fig.~\ref{fig:scale} the NLO scale dependence of the 4FS calculation 
described in Sec.~\ref{sec:4F} is shown.
The scale dependence of the 5FS calculation is milder 
than that of the 4FS calculation.
The two calculations approach each other for scales smaller than 
$(m_{H^\pm}+m_t)$. 
Note that the choice of scale $\tilde{\mu}_f$
is not motivated by the argument illustrated in Fig.~\ref{fig:scale}, but the latter rather 
confirms the findings of the kinematical study that led to identify $\mu_f$ with $\tilde{\mu}_f$.

\begin{figure}
\begin{center}
	\includegraphics[width=0.48\textwidth]{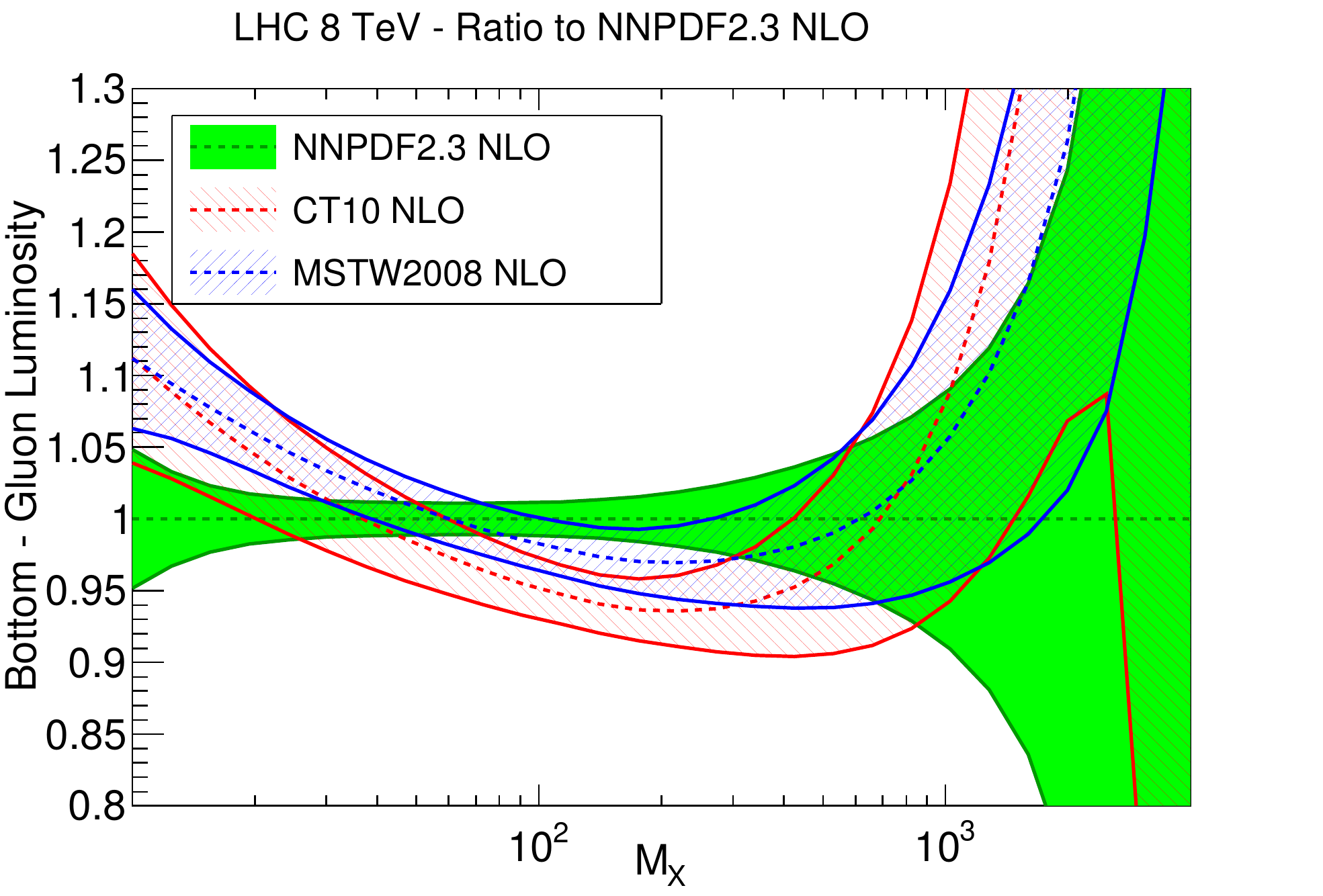}
	\includegraphics[width=0.48\textwidth]{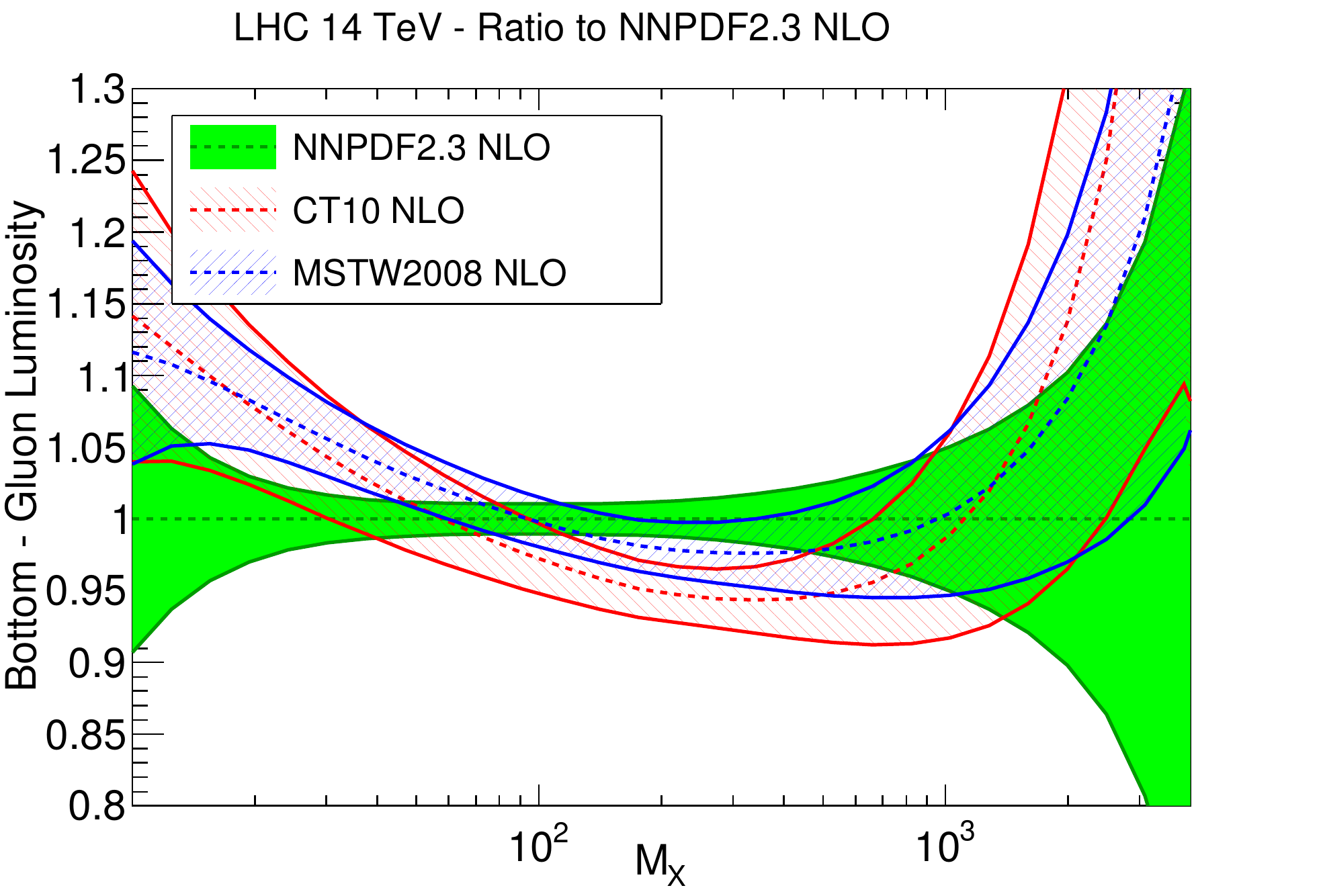}
\end{center}
	\caption{Gluon-bottom parton luminosities at the LHC for $\sqrt{s}=$ 8\,TeV (left) and 14\,TeV (right) 
for the default GM-VFN scheme NLO PDF sets CT10, MSTW2008 and NNPDF23, 
with $\alpha_s(M_Z)=0.118$ and $m_b= 4.75$\,GeV. Uncertainties correspond to 68\% C.L.}
	\label{5F:luminosities}
\end{figure}

In the 5FS computation the three GM-VFNS PDF sets mentioned in Sec.~\ref{sec:settings} are used:
the CT10 NLO set~\cite{CT10} and the corresponding set with $\alpha_s$ variation, 
the MSTW2008 NLO set~\cite{MSTW2008} and the corresponding sets with 
$\alpha_s$ and $m_b$ variations, the NNPDF2.3 NLO set~\cite{NNPDF23} 
and the corresponding sets needed to compute the $\alpha_s$ and $m_b$ variations. 
To illustrate the PDF uncertainty expected in the 5FS, the bottom-gluon luminosities for the three PDF sets, 
computed  with $\alpha_s(M_Z)=0.118$ and the default 
bottom quark mass $m_b= 4.75$\,GeV, are compared in Fig.~\ref{5F:luminosities} for the LHC at $\sqrt{s}=$ 8\,TeV and 14\,TeV. At a scale $M_X=m_{H^\pm}=200$\,GeV the 1$\sigma$ error bands of the NNPDF2.3 and CT10 
luminosities do not overlap, 
due to the harder gluon fitted by the NNPDF collaboration in the medium-to-large $x$ region.
At larger values of $m_{H^\pm}$ they tend to overlap, while at the same time 
the uncertainties become larger, driven by a larger gluon uncertainty at large values of $x$. 

For each PDF set $\alpha_s(M_Z)$ is varied by $0.0012$ around its central value~\cite{PDG}. 
The uncertainty due to the variation of $\alpha_s(M_Z)$ turns out to be negligible,
its size being about a order of magnitude smaller than the PDF uncertainty. 
This is expected since the gluon and the bottom quark have opposite-sign correlation 
with the value of $\alpha_s$ in the region of $x$ which is relevant for this process.
Therefore a partial cancellation of the $\alpha_s$ dependence is expected.

The uncertainty associated to the value of the bottom quark pole mass 
used in the PDF fits is estimated
by varying $m_b$ in the range $m_b$ = 4.75 $\pm$ 0.06\,GeV
\footnote{Note that CT10 provides sets to compute the PDF+$\alpha_s$ 
uncertainty, but does not provide sets associated to the $m_b$ variation. 
Therefore, in contrast to MSTW2008 and NNPDF2.3, the CT10 uncertainty is underestimated.}. 
In contrast to the $\alpha_s$ uncertainty, the $m_b$ uncertainty induced by the PDFs is quite significant, 
corresponding to about 30-50\% of the PDF uncertainty at fixed $m_b$. 
A significant dependence of the predicted cross section on the bottom-quark mass has already
been observed in several studies of processes initiated by bottom 
quarks~\cite{Bagnaschi:2014zla,NNPDF21,MSTW2008nf4}
and has to be taken into account for a realistic estimate of the theoretical uncertainty.

Results are summarized in Table~\ref{tab:5F:pdf}.
\begin{table}
\renewcommand*{\arraystretch}{1.35}
\begin{center}
  \begin{tabular}{|c|c|l|r|r|r| }
    \hline
 $\sqrt{s}$ [TeV], PDF set &    $m_{H^\pm}$ [GeV] & $\sigma_{\rm NLO}$ [pb] & $\delta_{\rm PDF}$ [\%] & $\delta\alpha_s$ [\%] & $\delta m_b$ [\%]\\
\hline
\hline
8    & 200 &  0.189        & $^{+6.7}_{-6.0}$   & $^{+0.6}_{-0.6}$      & n.a. \\
CT10 & 400 &  0.0289       & $^{+10.3}_{-\;\; 8.9}$  & $^{+1.2}_{-1.2}$      & n.a. \\
     & 600 &  0.00618      & $^{+14.7}_{-11.7}$ & $^{+1.9}_{-1.6}$  & n.a. \\
\hline
8        & 200 &  0.195    & $^{+4.7}_{-5.0}$  & $^{+0.5}_{-0.4}$ & $^{+2.8}_{-2.6}$ \\
MSTW2008 & 400 &  0.0292   & $^{+5.8}_{-7.9}$  & $^{+0.7}_{-0.8}$ & $^{+2.8}_{-2.7}$ \\
         & 600 &  0.00609  & $^{+8.3}_{-9.4}$  & $^{+1.3}_{-1.2}$  & $^{+2.9}_{-2.7}$ \\
\hline
8        & 200 &  0.195    & $\pm 4.4$   & $\pm 0.7$   & $\pm 1.2$ \\
NNPDF2.3 & 400 &  0.0288   & $\pm 6.6$   & $\pm 0.6$   & $\pm 2.8$ \\
         & 600 &  0.00586  & $\pm 8.9$   & $\pm 0.7$   & $\pm 1.5$ \\
\hline
\hline
14       & 200 & 0.870     & $^{+3.9}_{-3.5}$   & $^{+0.0}_{-0.0}$      & n.a. \\
CT10     & 400 & 0.171     & $^{+5.7}_{-5.2}$   & $^{+0.0}_{-0.0}$      & n.a. \\
         & 600 & 0.0458    & $^{+7.6}_{-6.9}$   & $^{+0.5}_{-0.5}$      & n.a. \\
\hline
14       & 200 & 0.902     & $^{+2.7}_{-3.6}$  & $^{+0.1}_{-0.0}$ & $^{+2.9}_{-2.7}$\\
MSTW2008 & 400 & 0.176     & $^{+4.0}_{-3.9}$  & $^{+0.0}_{-0.0}$ & $^{+2.9}_{-2.6}$\\
         & 600 & 0.0468    & $^{+4.7}_{-6.1}$  & $^{+0.0}_{-0.2}$ & $^{+2.9}_{-2.7}$\\
\hline
14       & 200 & 0.913     &  $\pm 2.7$   & $\pm 0.8$   & $\pm 1.5$\\
NNPDF2.3 & 400 & 0.179     &  $\pm 3.9$   & $\pm 0.6$   & $\pm 1.2$\\
         & 600 & 0.0471    &  $\pm 5.1$   & $\pm 0.5$   & $\pm 1.2$\\
\hline
  \end{tabular}
  \caption{\label{tab:5F:pdf}Central value and PDF, $\alpha_s$, $m_b$ uncertainty for the next-to-leading order
total $tH^-$ production cross section in the 5FS,
computed with different input PDF sets. Central values are computed with $\alpha_s(M_Z)=0.118$, $\alpha_s$ 
variation by varying $\alpha_s(M_Z)$ by $\pm 0.0012$ about its central value, $m_b$ variation by
varying the $m_b$ pole mass in the input PDFs by $\pm 60$ MeV. The n.a. in the boxes means that
there is no available PDF set to compute the corresponding variation.}
\end{center}
\end{table}
The predictions obtained with {\tt CT10} 
have the largest associated PDF uncertainty, 
as can be inferred from the luminosity
plots in Fig.~\ref{5F:luminosities}. 
Furthermore, the size of the PDF uncertainty increases with the
mass of the produced particles, the large-$x$ region being the one 
in which the gluon and the bottom quark PDFs are least constrained by data.
Compared to the PDF uncertainty, the relative $m_b$ uncertainty is more significant for light 
charged Higgs boson masses ($m_{H^\pm}=200$\,GeV), with larger values of $m_b$ 
corresponding to smaller cross sections.
Note that there is a partial cancellation between the bottom-quark mass 
dependence of the PDF and the Yukawa coupling which 
can be understood as follows: increasing the bottom-quark mass in PDF fits reduces the 
phase space available for the splitting of gluons into bottom quarks and thus reduces the bottom PDF 
(a similar suppression is induced by the explicit logarithms of the bottom quark mass 
which appear in the 4FS calculation). 
Increasing the bottom quark mass in the Yukawa coupling, on the other hand, 
increases the Yukawa coupling and thus the cross section normalization. 
For the overall bottom-quark mass uncertainty, $\delta m_b$, 
reported in Table~\ref{tab:5F:pdf}, the $m_b$ dependence due to the PDF 
and due to the Yukawa coupling therefore partially cancel. 

\begin{figure}[ht]
\begin{center}
	\includegraphics[width=0.48\textwidth]{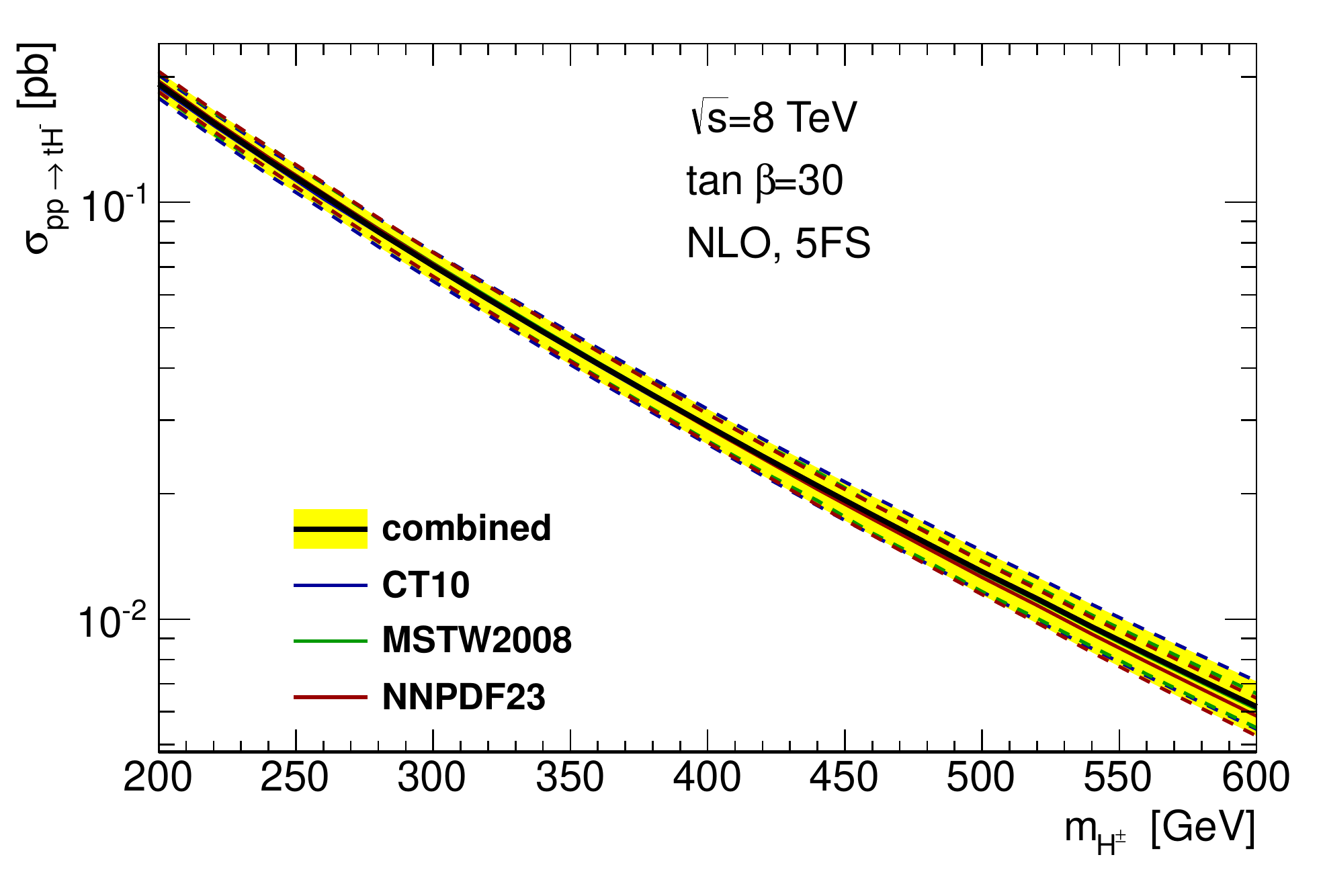}
	\includegraphics[width=0.48\textwidth]{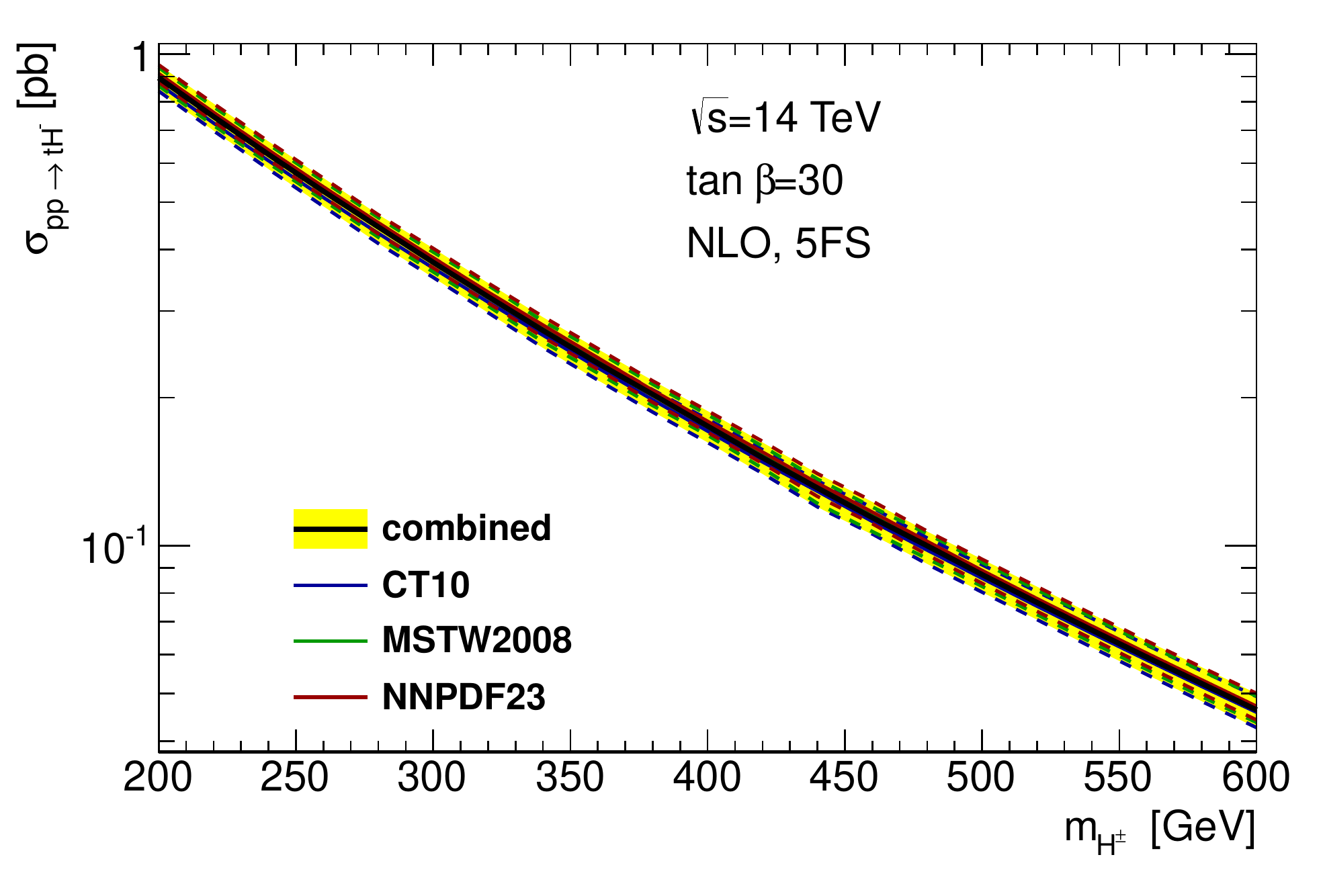}
\end{center}
	\caption{5FS cross section and PDF+$\alpha_s$+$m_b$ uncertainty for 
$pp \rightarrow tH^- + X$ at the LHC with $\sqrt{s}=$ 8 TeV (left) and 14\,TeV (right), 
calculated with CT10 (blue), 
MSTW2008 (green) and NNPDF2.3 (red) at NLO in a type-II 2HDM. The yellow
band corresponds to the envelope of the three predictions.}
	\label{5F:combined:plot}
\end{figure}

In Fig.~\ref{5F:combined:plot}, the predictions for the total cross section 
are presented for each of the three PDF sets. 
The error band corresponds to the total PDF+$\alpha_s$+$m_b$ uncertainty, computed
from the uncertainties shown in Table~\ref{tab:5F:pdf} according to Eq.~\eqref{eq:uncertainty}.
Moreover the combined prediction is presented, i.e. the envelope of the 
total PDF+$\alpha_s$+$m_b$ uncertainty of each prediction,
according to the PDF4LHC recommendation~\cite{LHCENVELOPE}, and as described in Sec.~\ref{sec:settings}. 
Taking the envelope significantly increases the size of the PDF uncertainty as obtained with each of the 
PDF sets individually, as it can also be inferred from Fig.~\ref{5F:luminosities}. 

Finally the scale uncertainty
due to missing higher orders, $\Delta^{\pm}_{\mu}$, obtained according to the prescription 
described in Sec.~\ref{sec:settings}, is linearly added to upper and lower bounds of the
envelope, according to Eq.~\eqref{eq:tot-uncertainty}. The variation of the renormalization scale 
and the scale in the Yukawa coupling contribute approximately equally to $\Delta^{\pm}_{\mu}$, 
while the impact 
of the factorization scale dependence is smaller by about a factor of 2. 
All individual sources of uncertainty, scale variation and total PDF uncertainty
(including $\alpha_s$ and $m_b$ variation) 
are listed in Table~\ref{tab:5F:unc}. 

\begin{figure}
\begin{center}
	\includegraphics[width=0.48\textwidth]{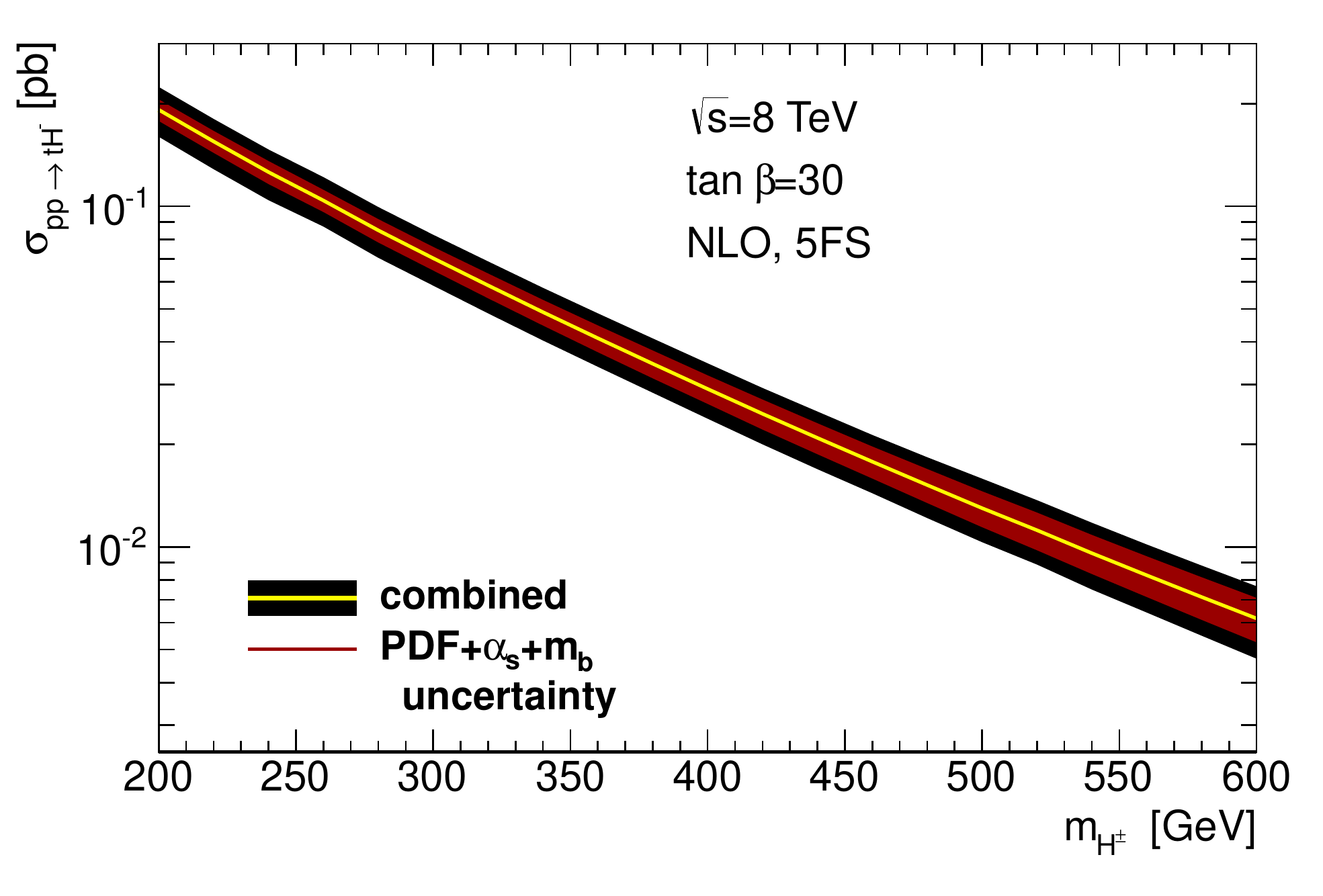}
	\includegraphics[width=0.48\textwidth]{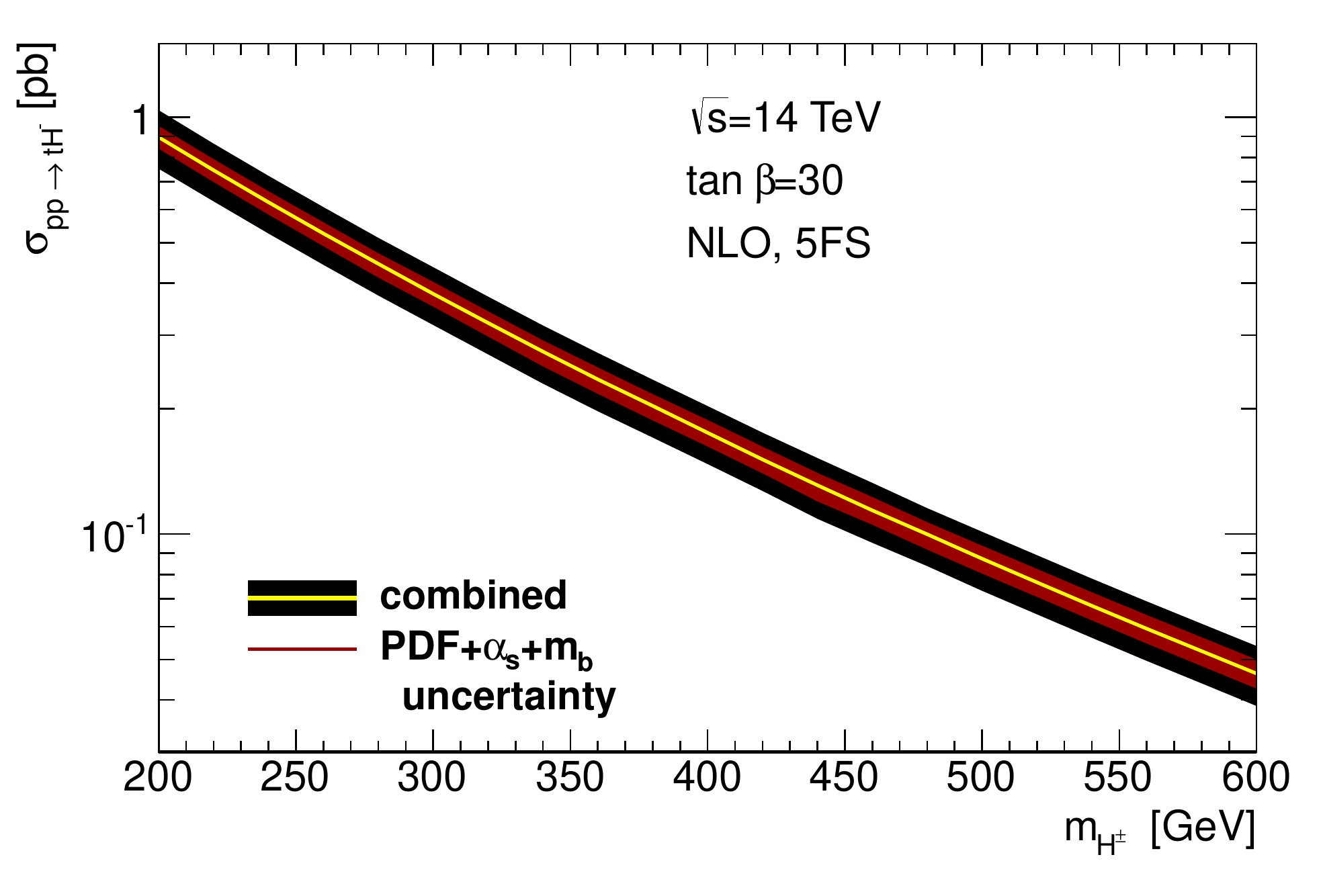}
\end{center}
	\caption{5FS cross section and uncertainties for $pp \rightarrow tH^- + X$ for 
the LHC at $\sqrt{s}=8$\,TeV (left) and 14\,TeV (right). Shown is the combined central value 
and the total uncertainty, split up into PDF+$\alpha_s$+$m_b$ and scale uncertainties.}
	\label{5F:combined:plot2}
\end{figure}

Fig.~\ref{5F:combined:plot2} displays the combined cross section, with its
total theoretical uncertainty 
split up into PDF+$\alpha_s$+$m_b$ uncertainty and scale variation. 
For the charged Higgs boson mass range considered, 
the size of the two uncertainties is comparable, with the total PDF uncertainty
being larger for higher $m_{H^\pm}$ masses.
\begin{table}
\renewcommand*{\arraystretch}{1.35}
\begin{center}
  \begin{tabular}{|c|c|c|c|c|c| }
    \hline
 $\sqrt{s}$ {\small [TeV]} &    $m_{H^\pm}$ {\small[GeV]} & $\sigma_{\rm NLO}$ [pb] & $\Delta^{\pm}_{\mu}$ [\%] & $\delta^{\pm}_{{\rm PDF}+\alpha_s+m_b}$ [\%] 
& $\Delta^{\pm}_{\rm tot}$ [\%]\\
\hline
8  & 200 &  0.192      & $^{+9.4}_{-9.4}$  & ${\pm 7.3}$  & $^{+16.7}_{-16.7}$  \\
   & 400 &  0.0291     & $^{+9.3}_{-8.6}$  & ${\pm 9.6}$  & $^{+18.9}_{-18.2}$  \\
   & 600 &  0.00617    & $^{+9.4}_{-8.9}$  & ${\pm 14.9}$ & $^{+24.3}_{-23.8}$  \\
\hline
14 & 200 & 0.895      & $^{+9.8}_{-9.7}$   & ${\pm 6.3}$  & $^{+16.1}_{-16.0}$  \\
   & 400 & 0.175      & $^{+8.6}_{-8.6}$   & ${\pm 7.3}$  & $^{+15.9}_{-15.9}$  \\
   & 600 & 0.0463     & $^{+8.4}_{-8.4}$   & ${\pm 8.0}$  & $^{+16.4}_{-16.4}$  \\
\hline
  \end{tabular}
  \caption{\label{tab:5F:unc}Central prediction, scale uncertainty, 
PDF+$\alpha_s$+$m_b$ uncertainty,
and total theoretical uncertainty for the next-to-leading order
$tH^-$ production cross section in the 5FS.}
\end{center}
\end{table}

%%%%%%%%%%%%%%%%%%%%%%%%%%%%%%%%%%%%%%%%%%%%%%%%%%%%%%%%%%%%%%%%

\section{Four-flavour scheme results}
\label{sec:4F}

In the four-flavor scheme, bottom quarks are created perturbatively in the hard part of the process, and 
the bottom quark mass is included exactly at all orders.
At leading order the partonic processes are given by
\begin{equation*}
\begin{array}{lcl}
\label{4F:lo_process}
q\bar{q},gg & \longrightarrow & t H^- \bar{b},
\end{array}
\end{equation*} 
while next-to-leading-order QCD corrections consists of virtual one-loop diagrams, 
gluon radiation and gluon (anti-)quark scattering~\cite{MKRAEMER}. 
The total theoretical uncertainty of the 4FS prediction is estimated according to 
Eq.~\eqref{eq:tot-uncertainty}. The dependence of the NLO cross section on the factorization 
and renormalization scales is illustrated in Fig.~\ref{fig:scale}. 
The scale uncertainty estimate is obtained by varying $\mu_r$, $\mu_f$ and 
$\mu_b$ by a factor two about their central values $\mu = \left(m_{H^\pm} + m_t + m_b\right)/3$, as 
described in Sec.~\ref{sec:settings}.

In the 4FS cross section calculation the fixed-flavour-number PDF sets 
with $n_f=4$ provided by the global PDF fitting collaborations are used. 
\begin{figure}
\begin{center}
	\includegraphics[width=0.48\textwidth]{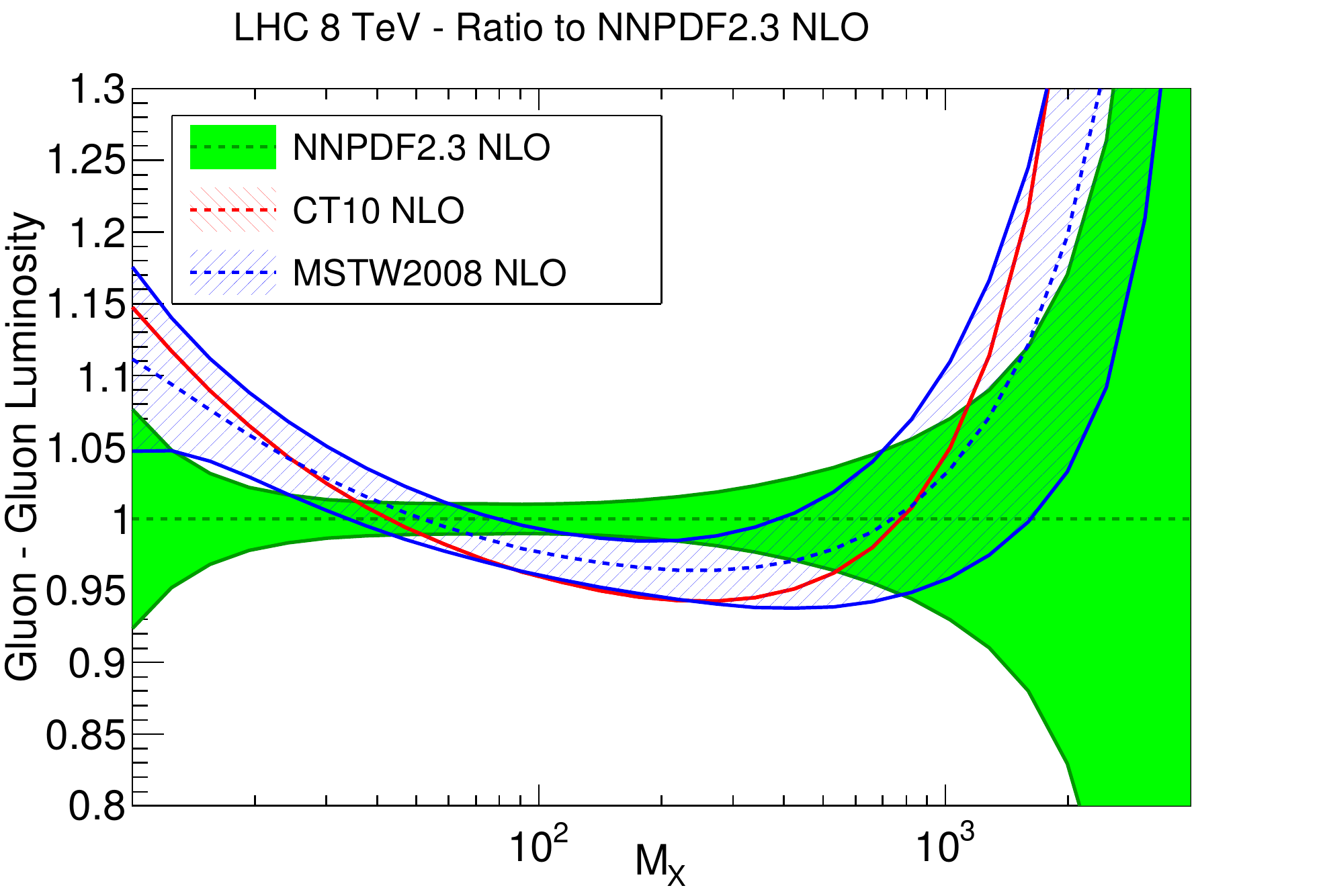}
	\includegraphics[width=0.48\textwidth]{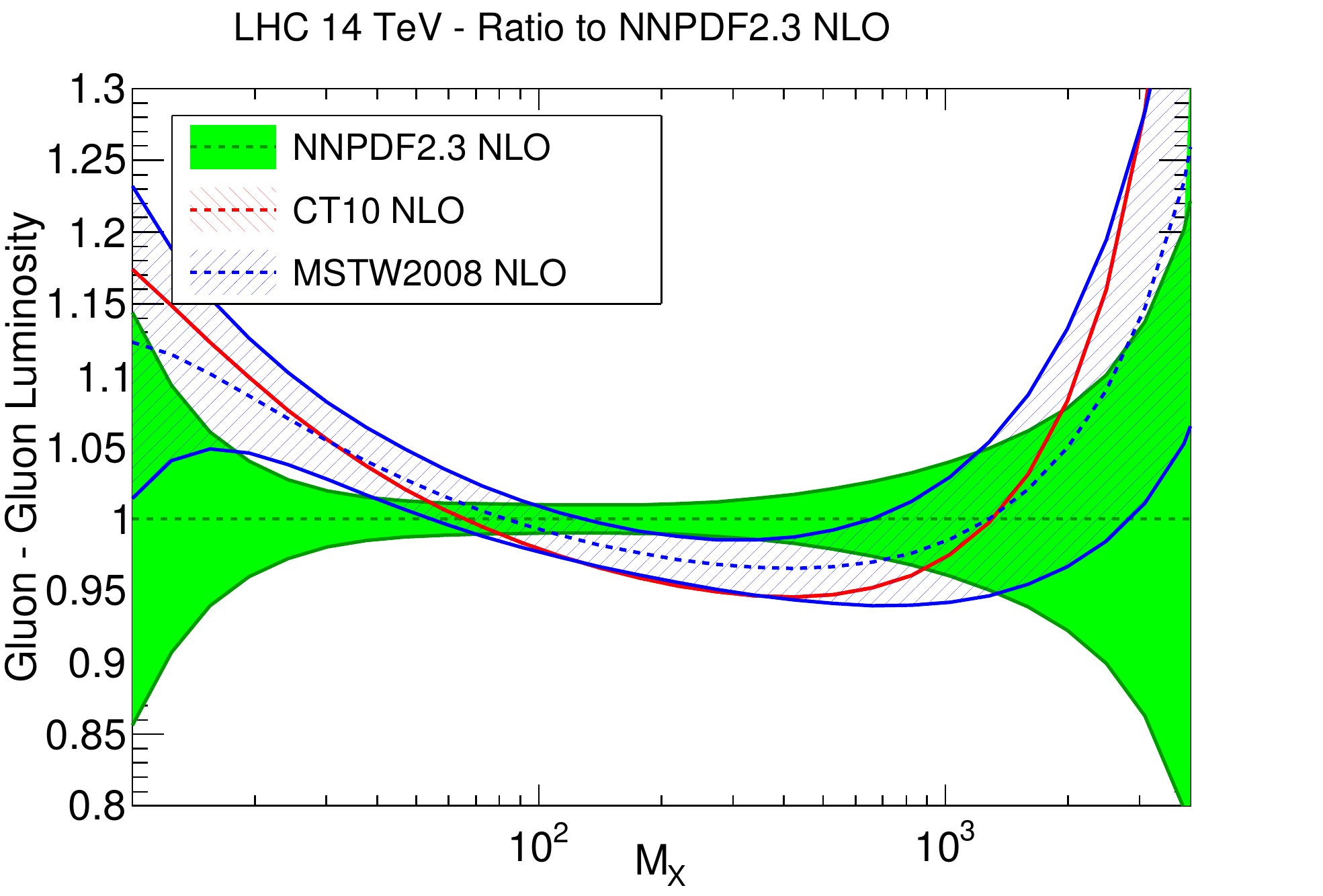}
\end{center}
	\caption{Gluon-gluon parton luminosities at the LHC at $\sqrt{s}=8$\,TeV (left) and 14\,TeV (right) for the 
4F Fixed-Flavour-Number NLO PDF sets provided by CT10, MSTW2008 and NNPDF2.3, 
with the default $\alpha_s^{4F}$ and $m_b$ values provided
by each collaboration. Uncertainties are at 68\% C.L.}
	\label{4F:luminosities}
\end{figure} 
The corresponding gluon-gluon luminosities are shown in Fig.~\ref{4F:luminosities}.  
Here, it is not possible to evaluate the PDFs at a common value of $\alpha_s$ since 
only the NNPDF collaboration provides FFN sets computed at 
various values of $\alpha_s^{4F}(M_z)$, while the MSTW and the CT collaborations only
provide 4FS PDF sets at their default $\alpha_s$ values, $\alpha_s^{4F}(M_z)= 0.1149$ for
the MSTW2008 NLO set and
$\alpha_s^{4F}(M_z) = 0.1127$ forthe 4FS CT10 NLO set.
Moreover CT10 do not provide an estimate of the PDF uncertainty for their $n_f=4$ set.
This leads to a slight underestimate in the total PDF uncertainty based on the
envelope of the three predictions.
In the charged Higgs boson mass range considered in this analysis
the NNPDF2.3 and the MSTW2008 gluon-gluon luminosities barely overlap within
their $1\sigma$ error bar and the CT10 central curve lies at the bottom of the
MSTW curve. This behaviour reflects the features observed  in 
Fig.~\ref{5F:luminosities}, as a consequence of the correlation between
the bottom and the gluon PDFs.

Since the 4FS MSTW2008 NLO sets are provided at several values of $m_b$ 
it is possible to estimate the PDF+$m_b$ uncertainty for the 4FS predictions
by varying $m_b$ in the input PDF sets, in the hard matrix element and in $m_b(m_b)$ 
of the bottom Yukawa coupling. The variation of the pole $m_b$ mass in a 4FS scheme 
PDF set is expected to have little influence, 
as $m_b$ only enters the partonic DIS cross section fitted in the global fit of 
PDFs.
The major dependence on $m_b$ comes from the variation of $m_b$ in the hard 
matrix element through the collinear logarithms
and from its variation in the bottom Yukawa coupling. Note that in the four-, as well
as in the five-flavor scheme, a partial cancellation occurs between the bottom quark mass dependence of the 
collinear logarithms and the Yukawa coupling. The latter
dominates in this case and the cross section increases as $m_b$ increases.

On the other hand, the NNPDF2.3 FFN sets are given at several values
of $\alpha_s$ and this allows one to estimate
the PDF+$\alpha_s$ uncertainty of the 4FS prediction. 
\begin{table}
\renewcommand*{\arraystretch}{1.35}
\begin{center}
  \begin{tabular}{|c|c|l|r|r|r| }
    \hline
 $\sqrt{s}$[TeV], PDF set &    $m_{H^\pm}$[GeV] & $\sigma_{\rm NLO}$ [pb] & $\delta_{\rm PDF}$ [\%] & $\delta\alpha_s$ [\%] & $\delta m_b$ [\%]\\
\hline
\hline
8        & 200 &  0.198   & n.a  & n.a.  & n.a. \\
CT10     & 400 &  0.0293  & n.a  & n.a.  & n.a. \\
         & 600 &  0.00608 & n.a  & n.a.  & n.a. \\
\hline
8        & 200 &  0.205   & $^{+2.3}_{-2.7}$  & n.a. & $^{+2.5}_{-2.7}$ \\
MSTW2008 & 400 &  0.0299  & $^{+7.9}_{-7.6}$  & n.a. & $^{+2.4}_{-2.6}$ \\
         & 600 &  0.00610 & $^{+17.8}_{-20.8}$  & n.a. & $^{+2.5}_{-2.5}$ \\
\hline
8        & 200 &  0.203   & $\pm 4.3$   & $^{+1.1}_{-0.8}$ & n.a. \\
NNPDF2.3 & 400 &  0.0288  & $\pm 6.4$   & $^{+0.8}_{-1.2}$ & n.a. \\
         & 600 &  0.00569 & $\pm 8.4$   & $^{+0.8}_{-2.2}$ & n.a. \\
\hline
\hline
14       & 200 & 0.938    & n.a  & n.a.  & n.a. \\
CT10     & 400 & 0.180    & n.a. & n.a.  & n.a. \\
         & 600 & 0.0475   & n.a. & n.a.  & n.a. \\
\hline
14       & 200 & 0.972    & $^{+1.1}_{-0.8}$  & n.a. & $^{+2.5}_{-2.6}$\\
MSTW2008 & 400 & 0.186    & $^{+2.2}_{-2.7}$  & n.a. & $^{+2.6}_{-2.6}$\\
         & 600 & 0.0489   & $^{+6.9}_{-5.5}$  & n.a. & $^{+2.6}_{-2.5}$\\
\hline
14       & 200 & 0.983    &  $\pm 2.6$   & $^{+0.1}_{-0.0}$  & n.a.\\
NNPDF2.3 & 400 & 0.187    &  $\pm 3.8$   & $^{+0.4}_{-0.5}$  & n.a.\\
         & 600 & 0.0481   &  $\pm 5.0$   & $^{+0.6}_{-0.9}$  & n.a.\\
\hline
  \end{tabular}
  \caption{\label{tab:4F:pdf}Central value and PDF, $\alpha_s$, $m_b$ uncertainties 
for the next-to-leading order
total $tH^-$ production cross section in the 4FS,
computed with different input PDF sets. Central values are computed with 
the default value of $\alpha_s(M_Z)$ provided by each PDF set, $\alpha_s$ 
variation by varying $\alpha_s(M_Z)$ by $\pm 0.0012$ about its central value, $m_b$ variation by
varying the $m_b$ pole mass in the input PDFs and in the hard
matrix element by $\pm 60$ MeV. The n.a. in the boxes means that
there is no available PDF set to compute the corresponding variation.}
\end{center}
\end{table}
Results are collected in Table~\ref{tab:4F:pdf}, in which the
relative uncertainty for each contribution is reported.
The central value for the next-to-leading order
cross section computed in the 4F scheme predicted by using the CT10 set is always
smaller with respect to the predictions obtained with MSTW2008 or NNPDF2.3, partially because the 
$\alpha_s^{4F}$ of the CT10 fit is smaller than that of the other PDF sets.
Furthermore, the PDF uncertainty increases as the mass of the produced Higgs boson increases,
a consequence of the rise in the gluon uncertainty at large $x$. 
It turns out that the $\alpha_s$ uncertainty contributes
very little to the total uncertainty while the $m_b$ variation 
induces an additional constant uncertainty about 2.5\% in the
whole $m_{H^\pm}$ range of our theoretical predictions. The latter 
is non negligible compared to the PDF error band,
especially for lighter charged Higgs. 
\begin{figure}
\begin{center}
	\includegraphics[width=0.48\textwidth]{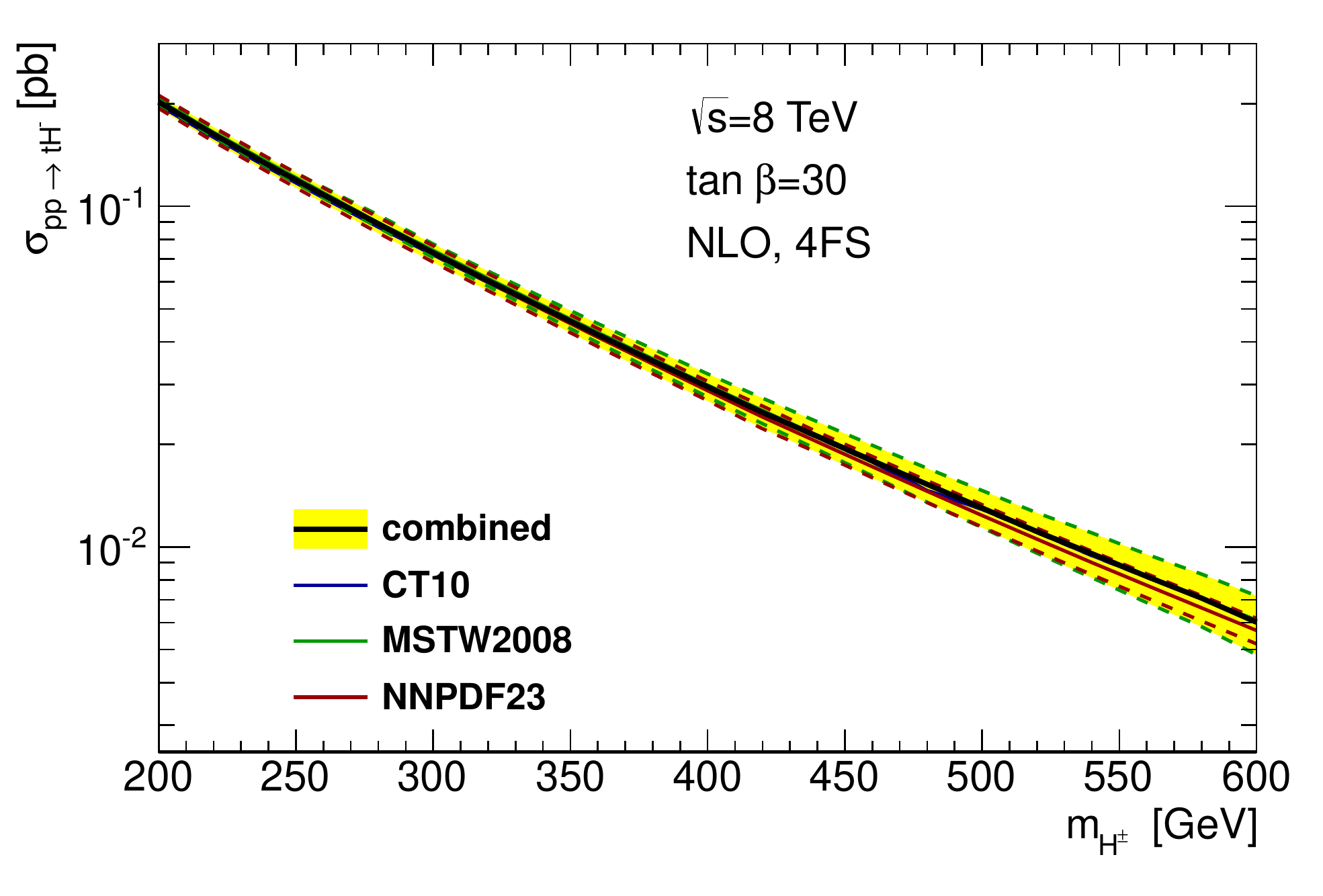}
	\includegraphics[width=0.48\textwidth]{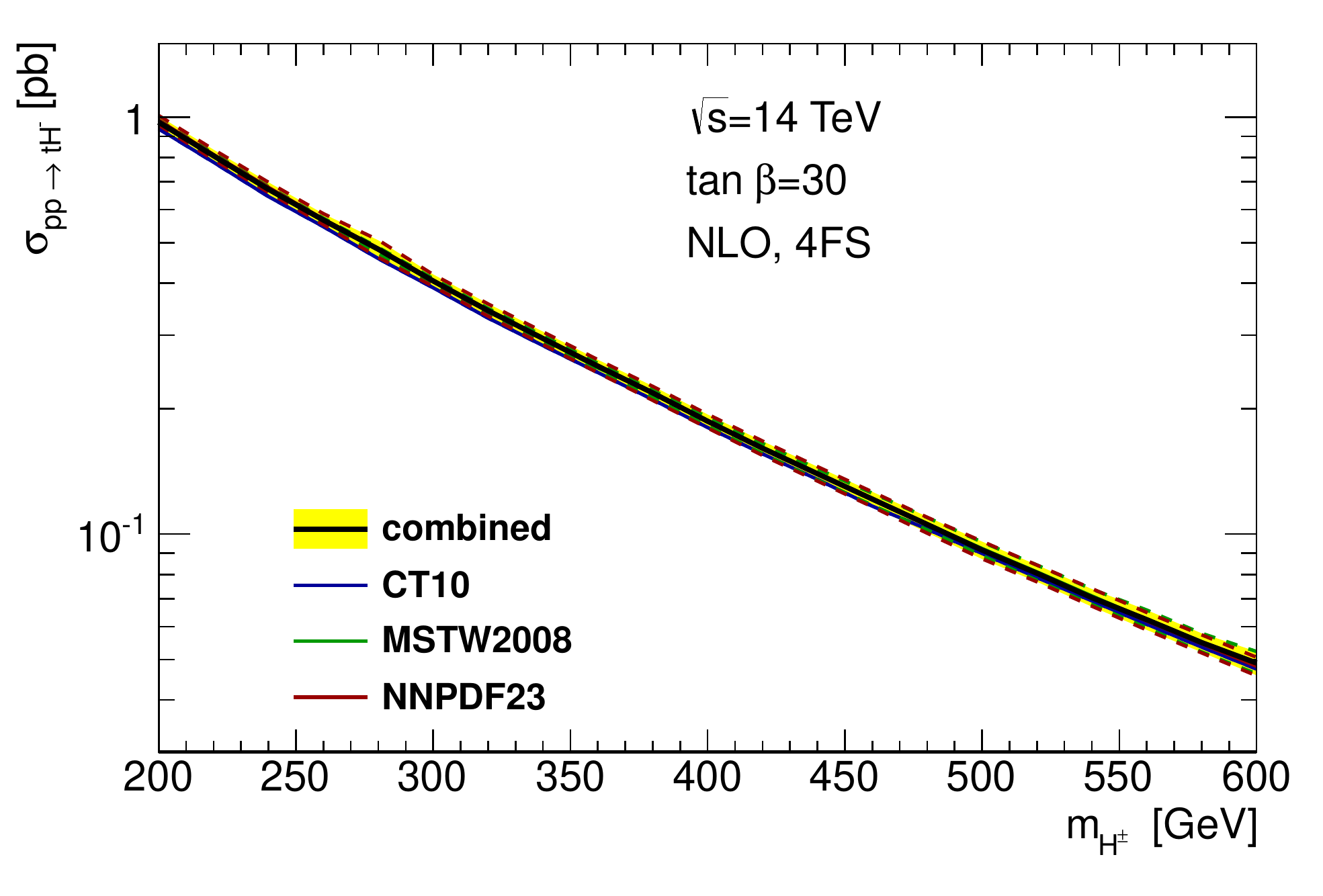}
\end{center}
        \caption{4FS cross section and PDF+$\alpha_s$+$m_b$ uncertainty for
$pp \rightarrow tH^- + X$ at the LHC with $\sqrt{s}=$ 8 and 14\,TeV , calculated with CT10 (blue),
MSTW2008 (green) and NNPDF2.3 (red) at NLO in a type-II 2HDM. The yellow
band corresponds to the envelope of the three predictions.
}
	\label{4F:combined:plot}
\end{figure} 

A comparison of the predicted cross section for the full range of charged Higgs boson masses
considered is shown in Fig.~\ref{4F:combined:plot}. 
The large uncertainty of the MSTW2008 gluon at large $x$ is at the origin 
of the large PDF uncertainty at large $m_{H^\pm}$ masses at $\sqrt{s}=8$\,TeV. At larger 
hadronic center-of-mass energies,
the gluon distributions peak at a smaller value of $x$ where the PDF uncertainty is less 
pronounced. Note that for large Higgs boson masses the average parton momentum fraction $x$, and thus the 
PDF uncertainty, is larger in the 4FS than in the 5FS. 

\begin{figure}
\begin{center}
	\includegraphics[width=0.48\textwidth]{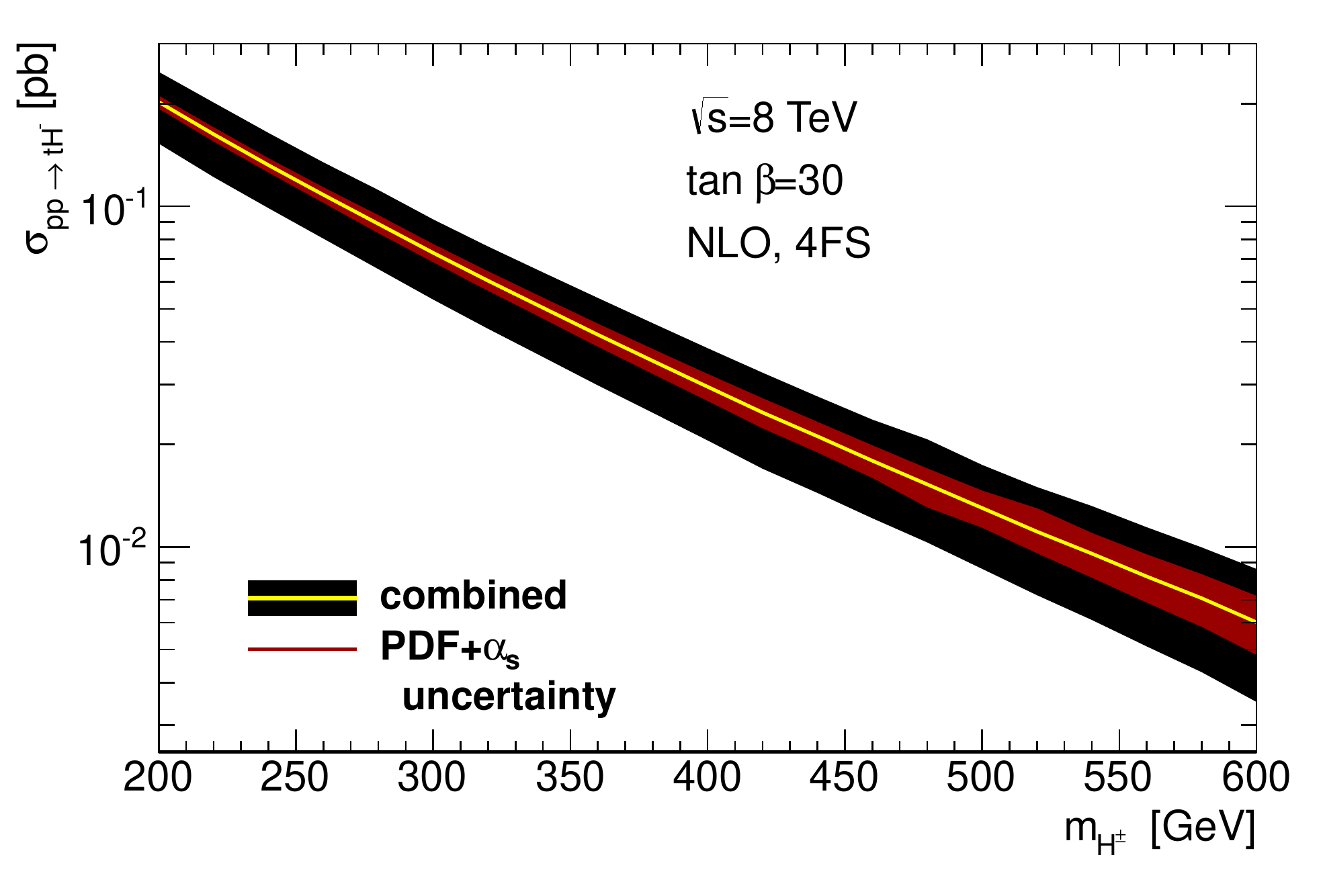}
	\includegraphics[width=0.48\textwidth]{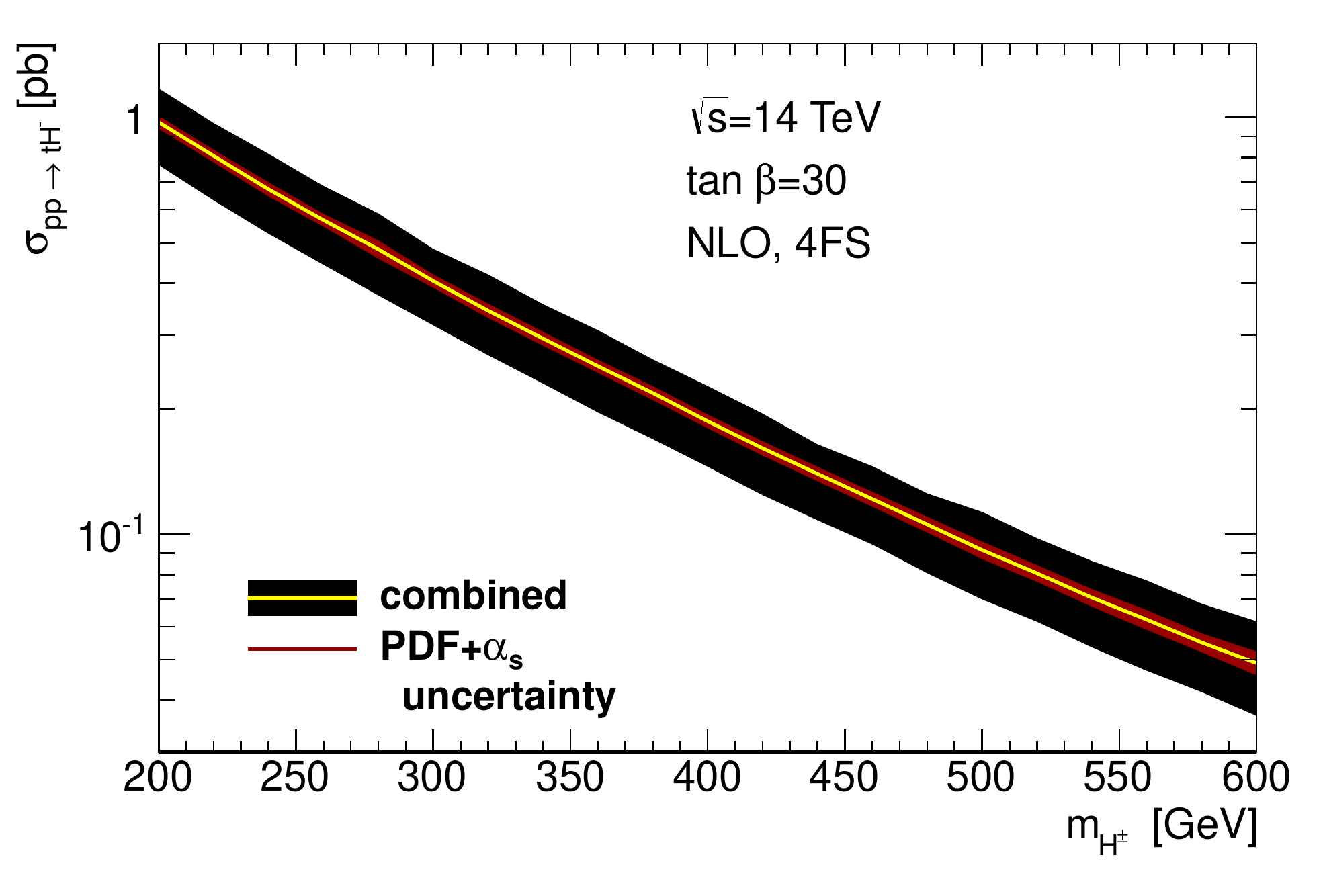}
\end{center}
	\caption{4FS cross section and uncertainties for $pp \rightarrow tH^- + X$ 
for the LHC at $\sqrt{s}=8$\,TeV (left) and 14\,TeV (right). 
Shown is the combined central value and the total uncertainty, 
split up into PDF+$\alpha_s$+$m_b$ and scale uncertainties.
}
	\label{4F:combined:plot2}
\end{figure}

Finally the uncertainty due to missing higher order contributions is linearly 
added to upper and lower bounds of the
envelope, according to Eq.~\eqref{eq:tot-uncertainty}.
Note that the scale in the running $m_b$ mass in the Yukawa coupling
contributes significantly to the scale variation, and contributes about 5 percentage points 
to the total scale uncertainty of 15\% to 20\% in the range of Higgs boson masses considered. 
Neglecting the $\mu_b$ variation would therefore lead
to an underestimate of the scale uncertainty.
All individual sources of uncertainty - scale variation, total PDF uncertainty
(including $\alpha_s$ and $m_b$ variation) and total uncertainty -
are listed in Table~\ref{tab:4F:unc}. 
The results in the table show that overall the scale uncertainty is  
the dominating source of theoretical
uncertainty for lower $m_{H^\pm}$ masses. The same
can be observed in Fig.~\ref{4F:combined:plot2} in which the central prediction 
for the 4F scheme cross section is presented, with its uncertainty split up into 
scale and PDF+$\alpha_s$+$m_b$ uncertainties.  
At large $m_{H^\pm}$ masses, especially at 8\,TeV,
the large-$x$ gluon uncertainty drives the total theoretical uncertainty 
above 40\%. The release of PDF sets that include
the constraints from the precise LHC jet and top data will help in reducing the
uncertainty of the gluons at large $x$ and consequently decrease the uncertainty
of the theoretical predictions.
\begin{table}
\renewcommand*{\arraystretch}{1.35}
\begin{center}
  \begin{tabular}{|c|c|c|c|c|c| }
    \hline
 $\sqrt{s}$[TeV] &    $m_{H^\pm}$[GeV] & $\sigma_{\rm NLO}$ [pb] & $\Delta^{\pm}_{\mu}$ [\%] & $\delta^{\pm}_{{\rm PDF}+\alpha_s+m_b}$ [\%] 
& $\Delta^{\pm}_{\rm tot}$ [\%]\\
\hline
8  & 200 &  0.203     & $^{+18.2}_{-20.1}$  & $\pm 4.6$  &   $^{+22.8}_{-24.7}$  \\   %$^{+6.2}_{-4.7}\qquad^{+0.4}_{-0.3}$ &
   & 400 &  0.0296    & $^{+20.6}_{-21.3}$  & $\pm 9.3$  &   $^{+29.9}_{-30.6}$  \\   %$^{+6.1}_{-4.8}\qquad^{+0.6}_{-0.5}$ &
   & 600 &  0.00601   & $^{+23.7}_{-22.0}$  & $\pm 19.8$ &   $^{+43.5}_{-41.8}$  \\   %$^{+6.1}_{-4.1}\qquad^{+0.7}_{-0.6}$ &
\hline
14 & 200 & 0.973     & $^{+16.8}_{-17.5}$   & $\pm 3.6$  &  $^{+20.4}_{-21.1}$  \\ %$^{+6.2}_{-4.0}\qquad^{+0.5}_{-0.5}$ &
   & 400 & 0.186     & $^{+17.7}_{-18.4}$   & $\pm 3.8$  &  $^{+21.5}_{-22.2}$  \\ %$^{+6.1}_{-3.9}\qquad^{+0.7}_{-0.6}$ &
   & 600 & 0.0492    & $^{+19.4}_{-18.7}$   & $\pm 6.8$  &  $^{+26.2}_{-25.5}$  \\ %$^{+6.0}_{-4.6}\qquad^{+0.7}_{-0.7}$ &
\hline
  \end{tabular}
  \caption{\label{tab:4F:unc}
 Central prediction, scale uncertainty, 
PDF+$\alpha_s$+$m_b$ uncertainty,
and total theoretical uncertainty for the next-to-leading order
$tH^-$ production cross section in the 4FS.}
\end{center}
\end{table}

%%%%%%%%%%%%%%%%%%%%%%%%%%%%%%%%%%%%%%%%%%%%%%%%%%%%%%%%%%%%%%%%%%
\section{Comparison and matching}
\label{sec:match}

The four- and five-flavor schemes only yield identical results for the $pp \rightarrow tH^\pm + X$  cross sections 
in an all-order calculation, as was shown for instance in Ref.~\cite{ACOT}. 
At finite order, the schemes include different contributions, since the perturbative expansion is 
ordered differently. 
Thus, the predictions within the two schemes can be used to cross-check results, 
and to estimate the impact of neglected contributions at higher order.

To obtain a unique theoretical prediction which can be confronted with experimental data, 
the four- and five-flavor schemes can be combined using a 
prescription called Santander-matching~\cite{SANTANDER}.
In the asymptotic limits $m_{H^\pm}/m_b \rightarrow 1$ and $m_{H^\pm}/m_b \rightarrow \infty$, 
the four- and five-flavor schemes, respectively, provide the unique description 
of the cross section. For realistic Higgs boson masses in the range from 200\,GeV to 600\,GeV 
considered here, both schemes contribute with a given finite weight which depends on the charged Higgs boson 
mass~\cite{SANTANDER}. The difference between the
two approaches is formally logarithmic, and thus the dependence of their relative importance on the
Higgs boson is determined by a logarithmic term, i.e.\
\begin{equation}
\myMathSigma{matched} =  \frac{\myMathSigma{4F} + w\myMathSigma{5F}}{1 + w} \,,
\end{equation}
with the weight $w$ defined as
\begin{equation}
\label{santander_eq}
w  =  \log \frac{m_{H^\pm}}{m_b} - 2\,.
\end{equation}
This yields a weight of 100\% for the 5FS cross section \mySigma{5F} in the 
limit of $m_{H^\pm}/m_b \rightarrow \infty$ as desired. 
A weight of 50\% is given to both cross sections for $m_{H^\pm}$ around 100\,GeV, 
to reflect the observation that predictions for both schemes agree well in this region. 
The theoretical uncertainties are combined as
\begin{equation}
\Delta \sigma_{\rm tot,matched}^{\pm}  =  \frac{\Delta \sigma_{\rm
tot,4F}^{\pm} + w \Delta \sigma_{\rm tot,5F}^{\pm}}{1 + w} \,.
\end{equation}
The Santander-matching scheme is a pragmatic and simple approach to derive a unique prediction from 
the four- and five-flavor scheme results, and not based on a thorough field-theoretic analysis. 
However, the Santander-matched 
results encompass the essential features of the two schemes. 
The corresponding matched predictions and uncertainty 
estimates are expected to be close to the true cross section, in particular as 
the four- and five-flavor scheme calculations for heavy charged Higgs boson production 
with the improved scale setting prescription are in good mutual agreement.

\begin{figure}
\begin{center}
  \includegraphics[width=0.48\textwidth]{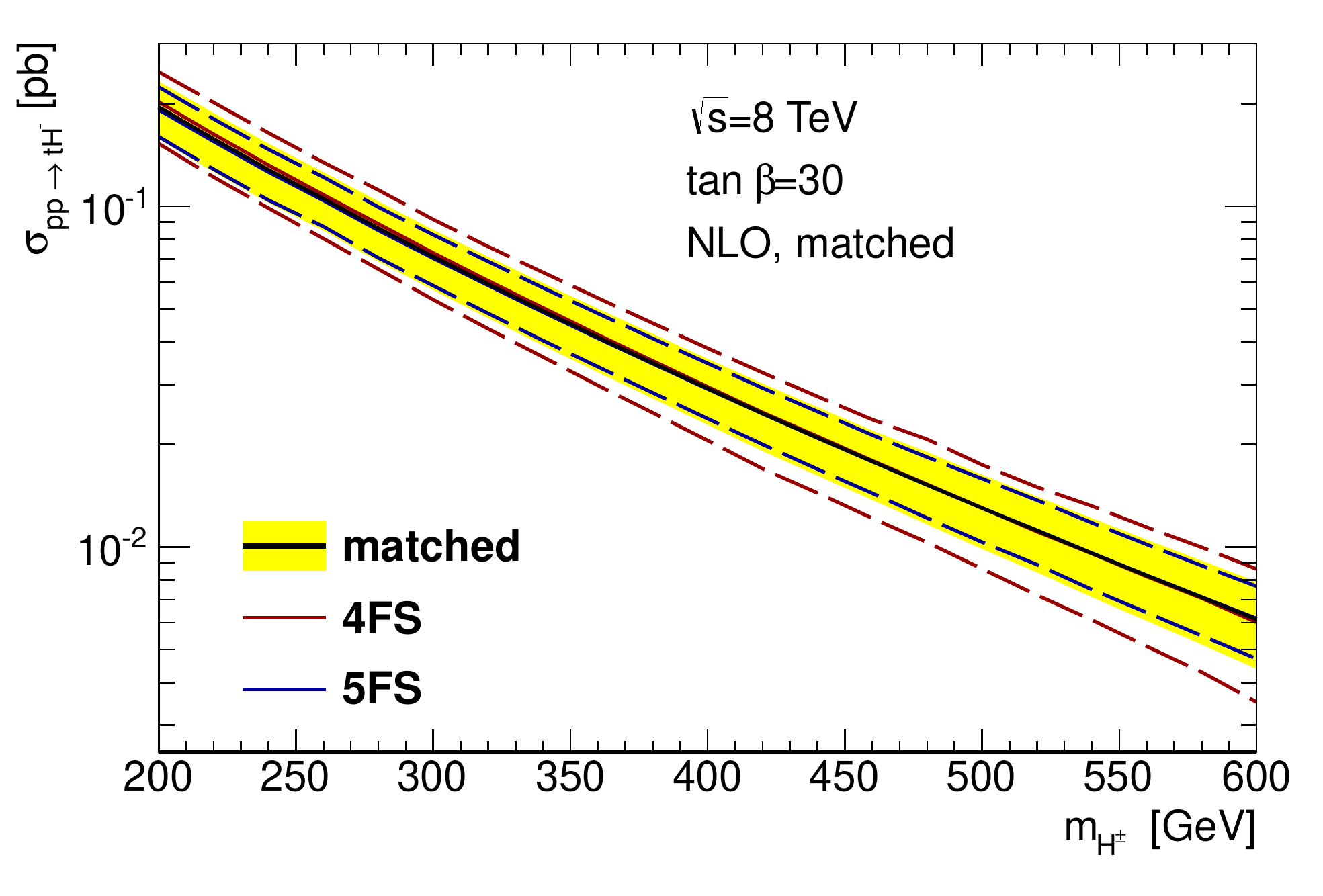}
  \includegraphics[width=0.48\textwidth]{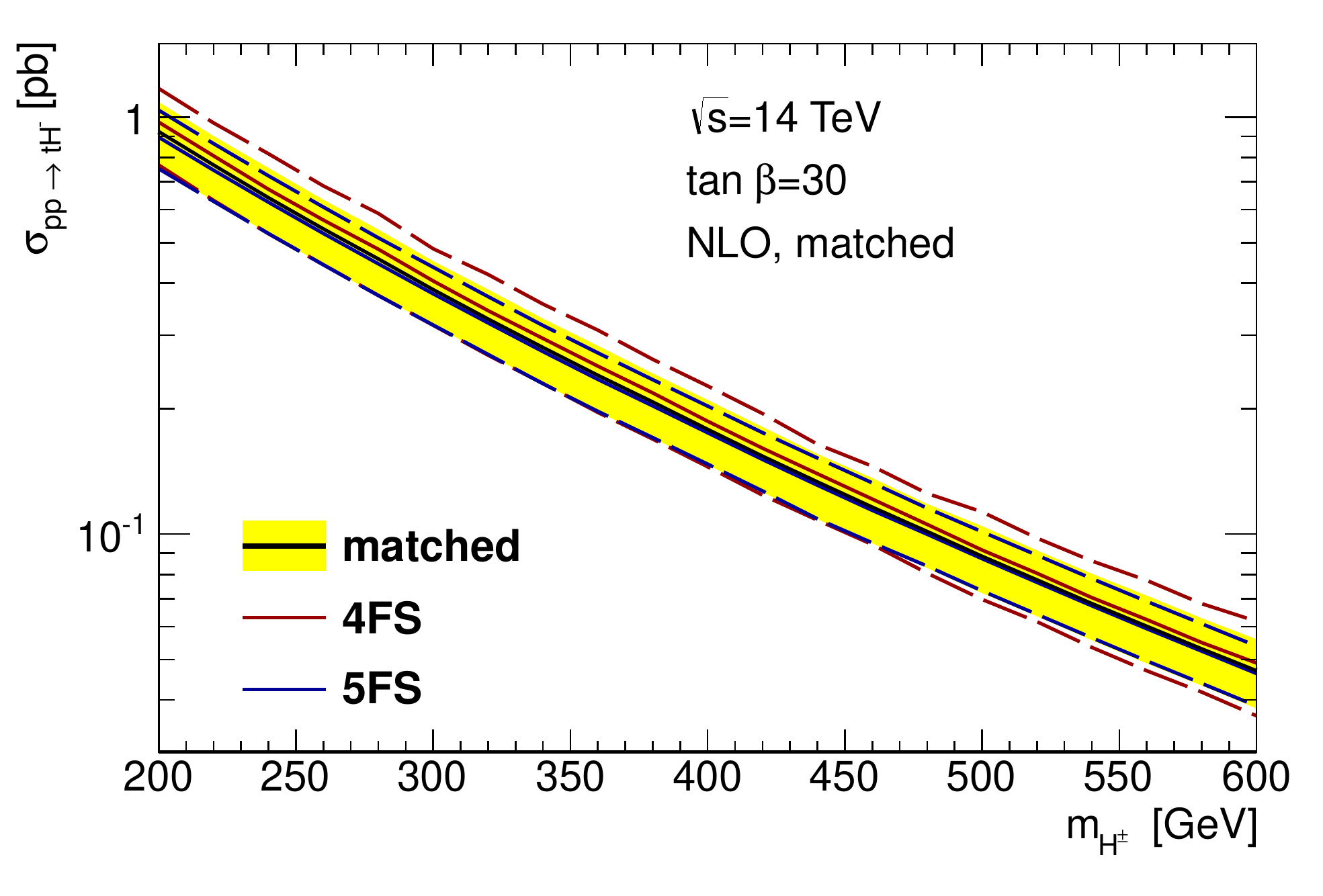}
\end{center}
        \caption{Santander-matched cross section and uncertainties for 
$pp \rightarrow tH^- + X$ at the LHC for 8 and 14\,TeV. 
The four- and five-flavor scheme results as well as the combined values are shown, 
together with their total uncertainties.
}
	\label{fig:matched}
\end{figure}

The cross section and uncertainty for the results of the four- and five-flavor scheme calculations and their combination 
for $\sqrt{s}=8$ and 14\,TeV are presented in Fig.~\ref{fig:matched}. 
The predictions from both schemes agree well within their uncertainties, with differences of at most 10\%. 
The prediction~\cite{MRU2012} that the impact of the resummation 
of the collinear logarithms decreases for higher masses of the produced heavy particle is confirmed. 
The overall theoretical uncertainty of the matched NLO prediction is about 20--30\%, 
very close to the 5F uncertainty that, for the considered range of masses, has a larger weight.

A much better agreement than in earlier comparisons~\cite{MKRAEMER} is observed. 
There, the choice for the factorization scale in the 5FS was $\mu_f$ = $\left(m_t + m_{H^\pm}\right)/3$. 
The dynamical choice for $\mu_f$ used here significantly improves the 
agreement between the predictions in the two schemes. In addition the
improved treatment of heavy quark threshold effects in the modern PDF sets employed here has lead
to a decrease of the bottom PDFs compared to previous analyses, and has thus moved the 5FS 
calculation closer to the 4FS cross section prediction. 

\section{Varying the parameter $\mathbf{\tan\beta}$}
\label{sec:tanbeta}
\begin{figure}
\begin{center}
  \includegraphics[width=0.48\textwidth]{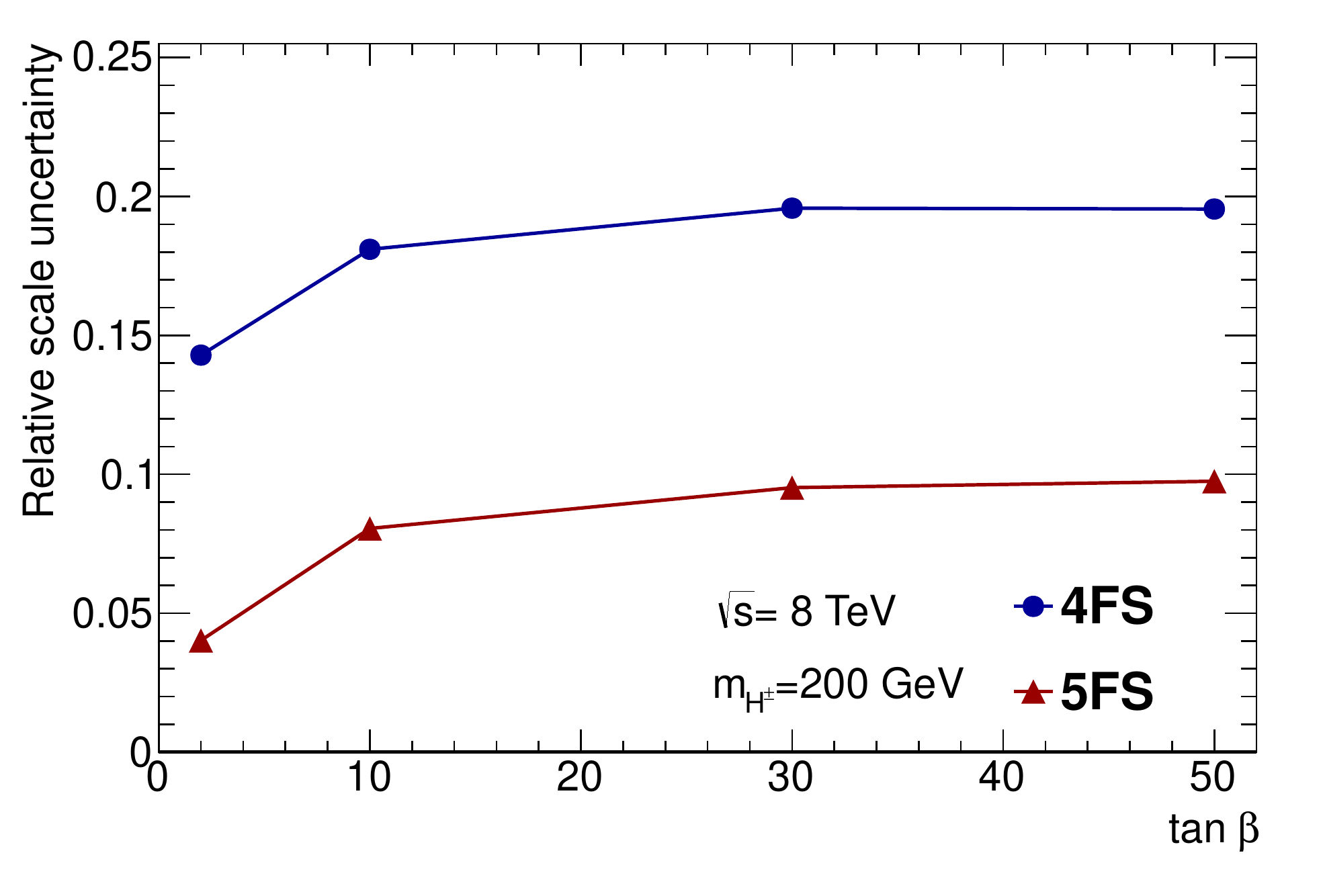}
  \includegraphics[width=0.48\textwidth]{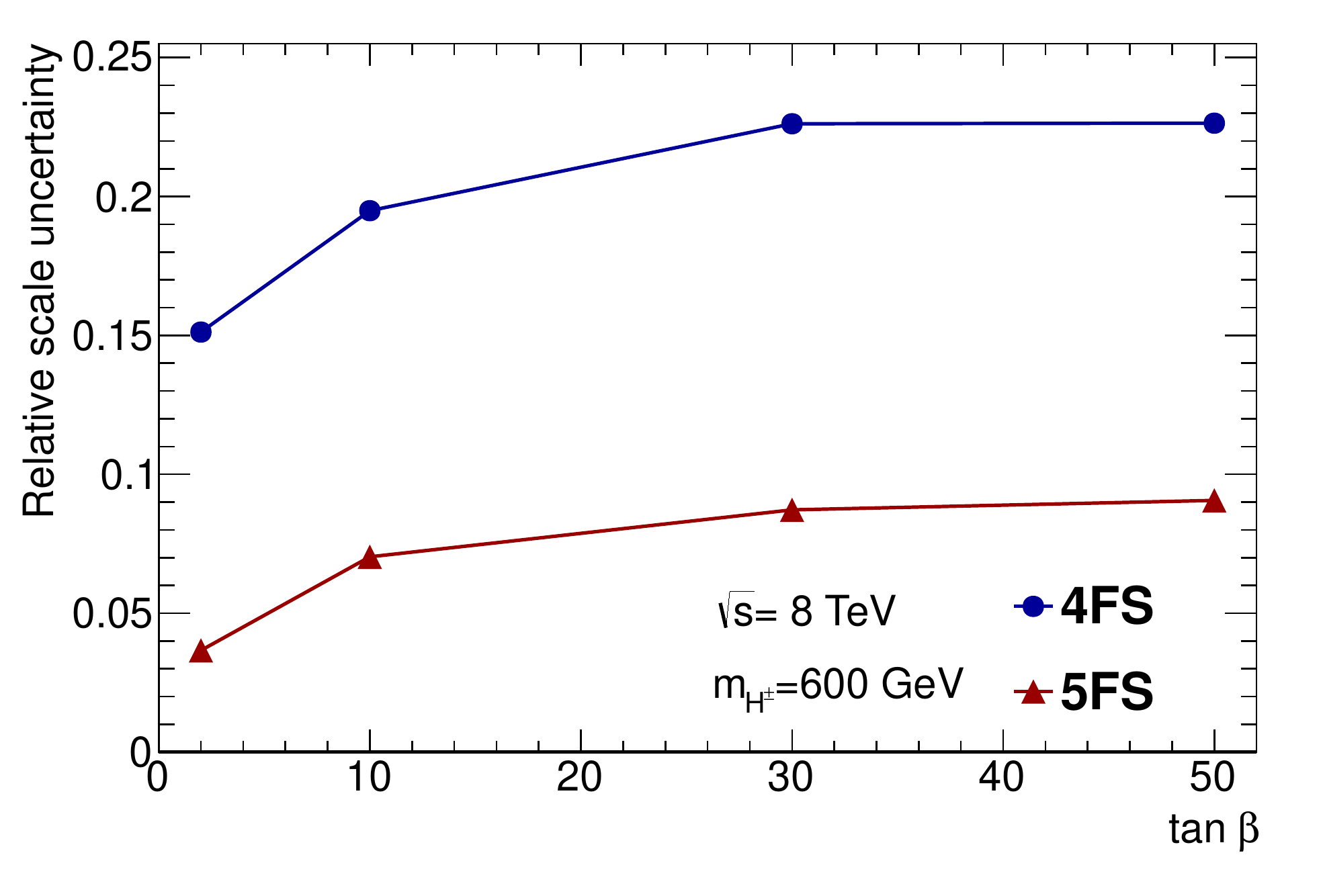}\\
  \includegraphics[width=0.48\textwidth]{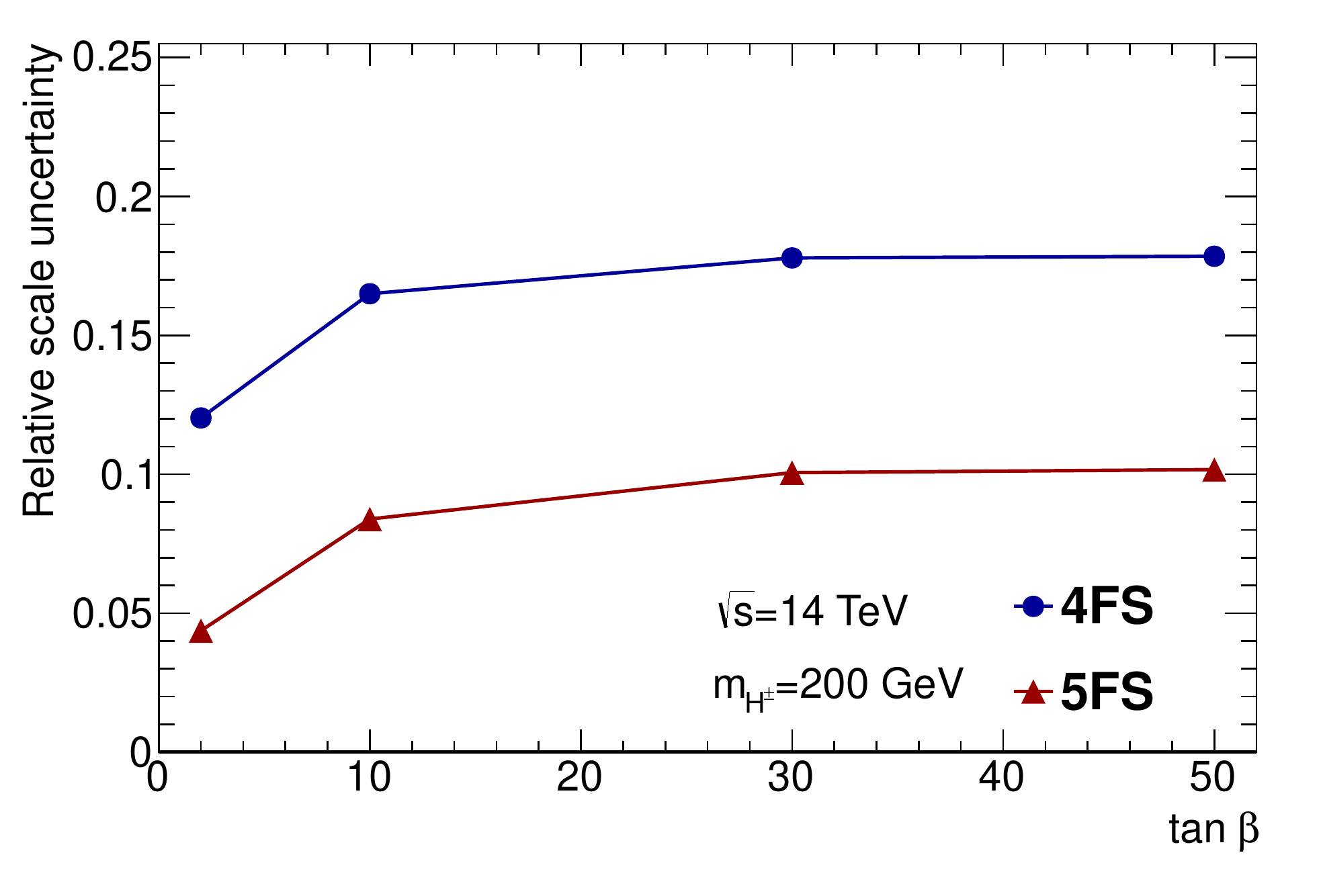}  
  \includegraphics[width=0.48\textwidth]{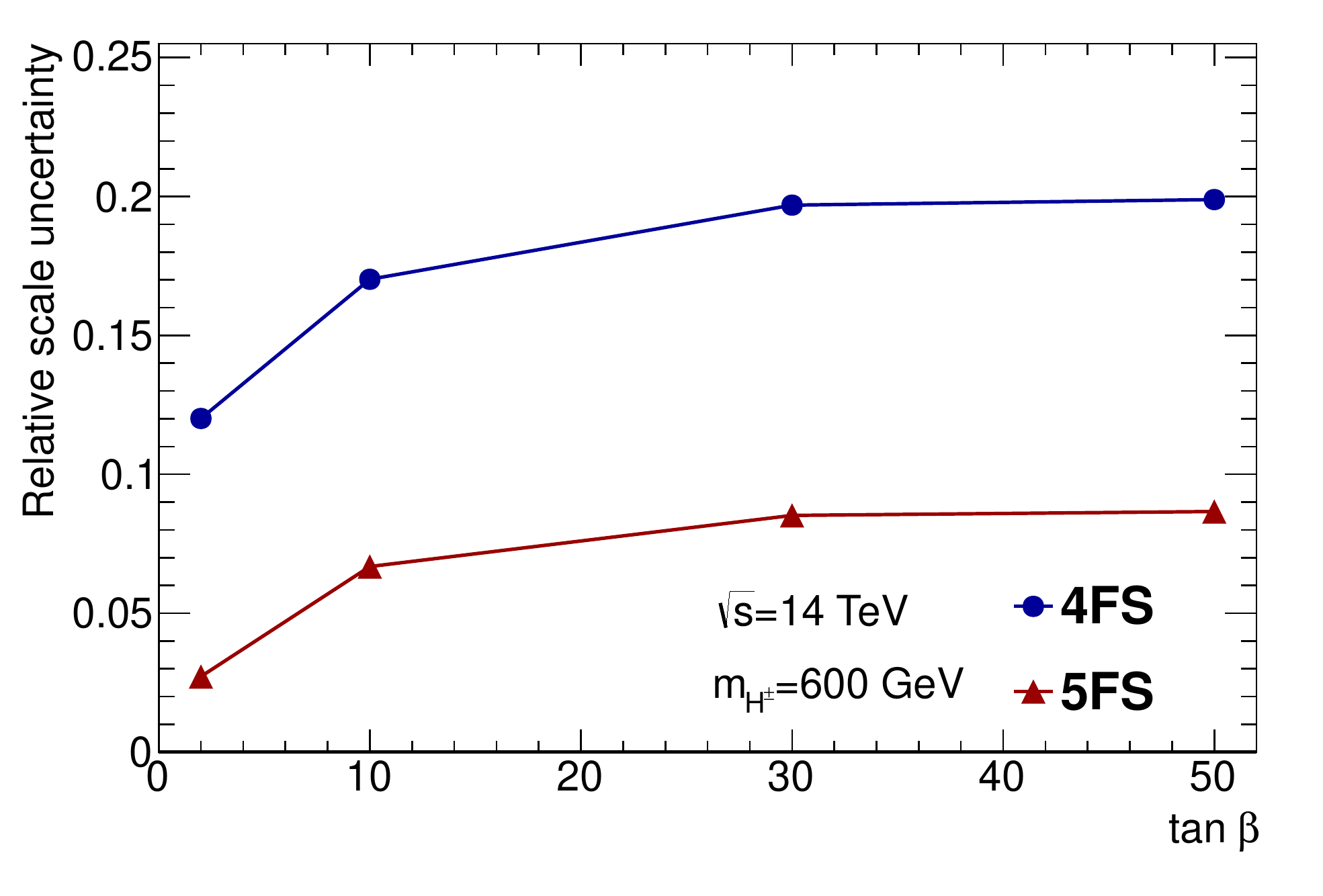}\\
\end{center}
        \caption{Scale dependence as a function of $\tan\beta$ for the 
4F and 5F predictions for the LHC at $\sqrt{s}=8$\,TeV (top row) and $\sqrt{s}=14$\,TeV (bottom row) for two 
values of $m_{H^\pm}$ masses: 200\,GeV (left) and 600\,GeV (right). 
}
        \label{fig:scaleunc}
\end{figure}

The cross section for charged Higgs boson production in association with a top quark and a bottom quark 
depends on the ratio of the vacuum expectation values of the two Higgs doublets, $\tan\beta = v_2/v_1$, 
through the Yukawa coupling, see Eq.~\eqref{eq:tbhcoupling}. The Yukawa coupling consists of two 
pieces which scale as $\tan\beta$ and $\cot\beta$, respectively. Thus changing $\tan\beta$ induces a non-trivial 
change in the cross section, but also in the theoretical uncertainty. First, the scale dependence 
as a function of $\tan\beta$ is considered for two values of the charged Higgs boson 
mass and the center-of-mass energies 8 and 14\,TeV in 
Fig.~\ref{fig:scaleunc}. A relatively uniform behavior is observed where the scale dependence decreases with decreasing 
$\tan\beta$ from about 20\% to 15\% and 10\% to 5\% for the four- and five-flavor scheme calculations, respectively. This 
is caused by the decreasing relevance of the running bottom Yukawa coupling, which 
is proportional to $\tan\beta$ and which adds about 5 percentage points to the overall scale uncertainty for large $\tan\beta$. 

\begin{figure}
\begin{center}
  \includegraphics[width=0.48\textwidth]{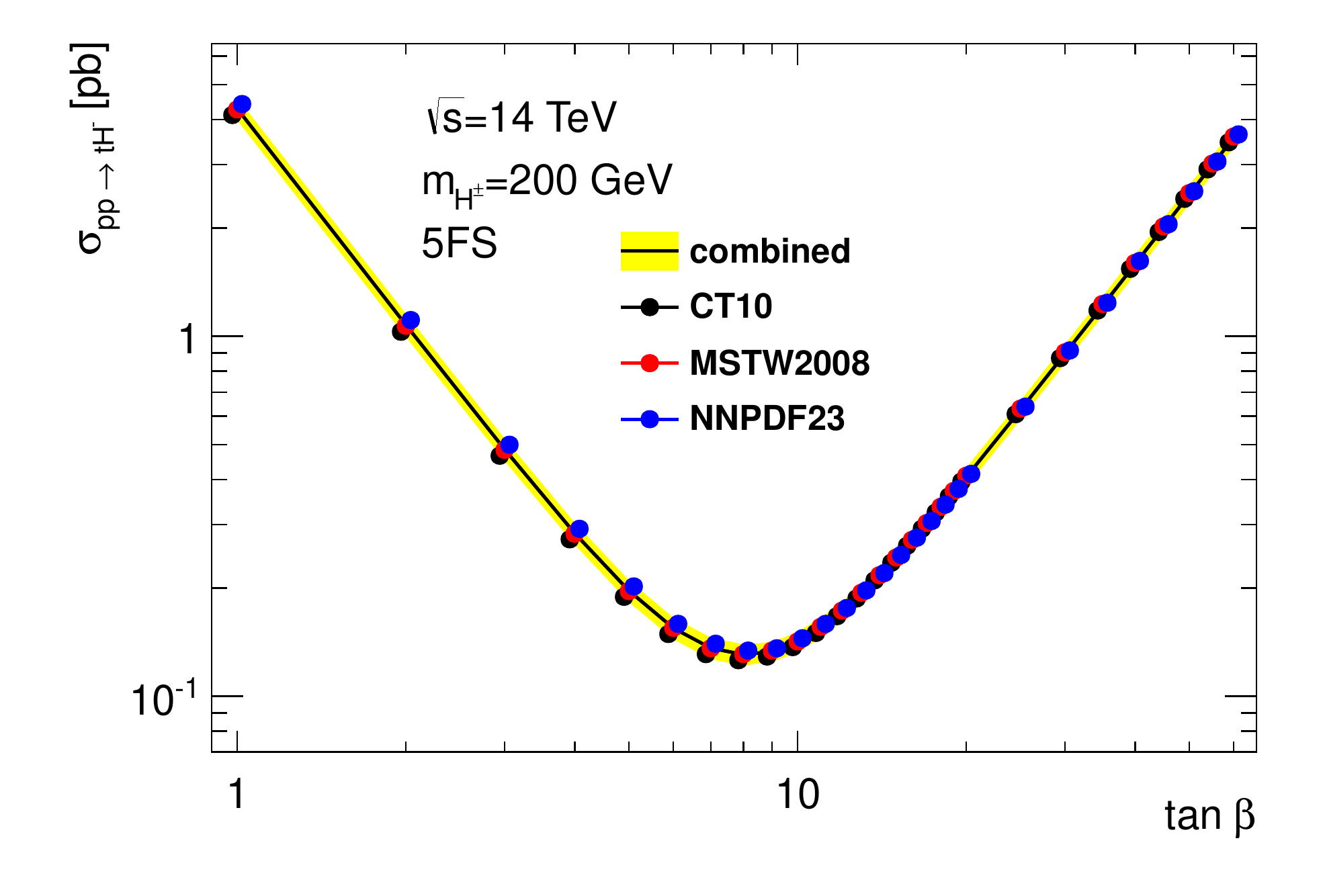}
  \includegraphics[width=0.48\textwidth]{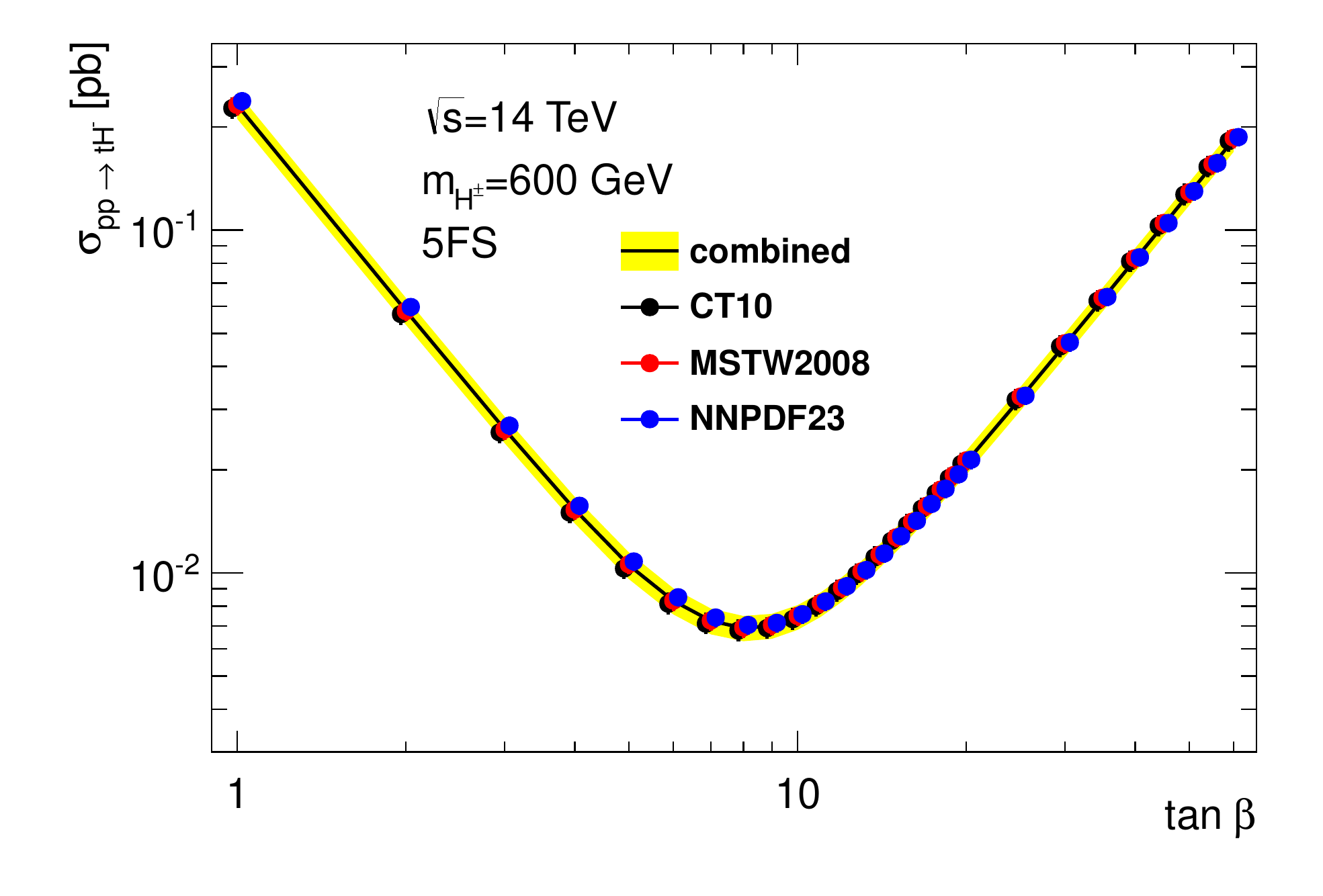}\\
  \includegraphics[width=0.48\textwidth]{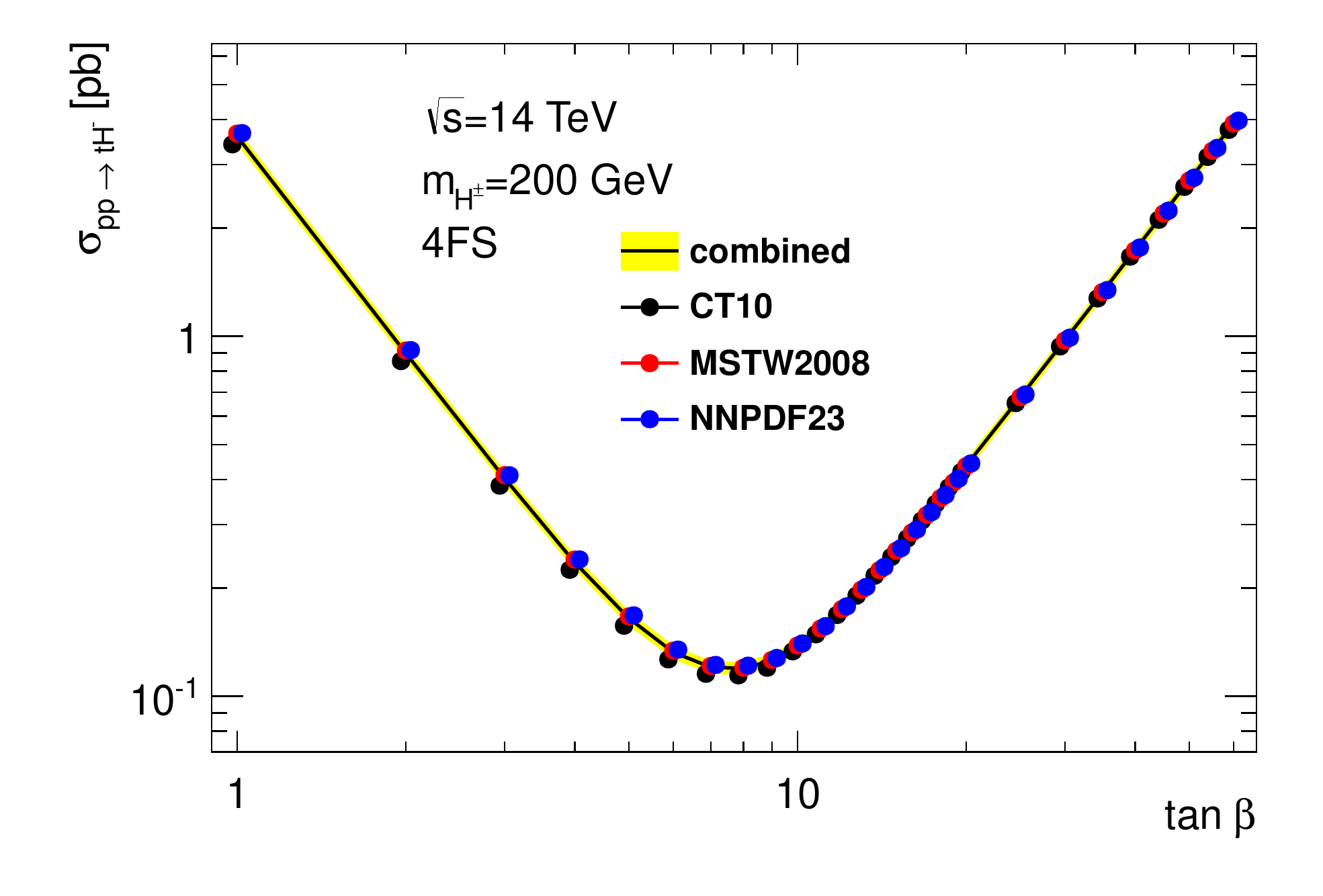}
  \includegraphics[width=0.48\textwidth]{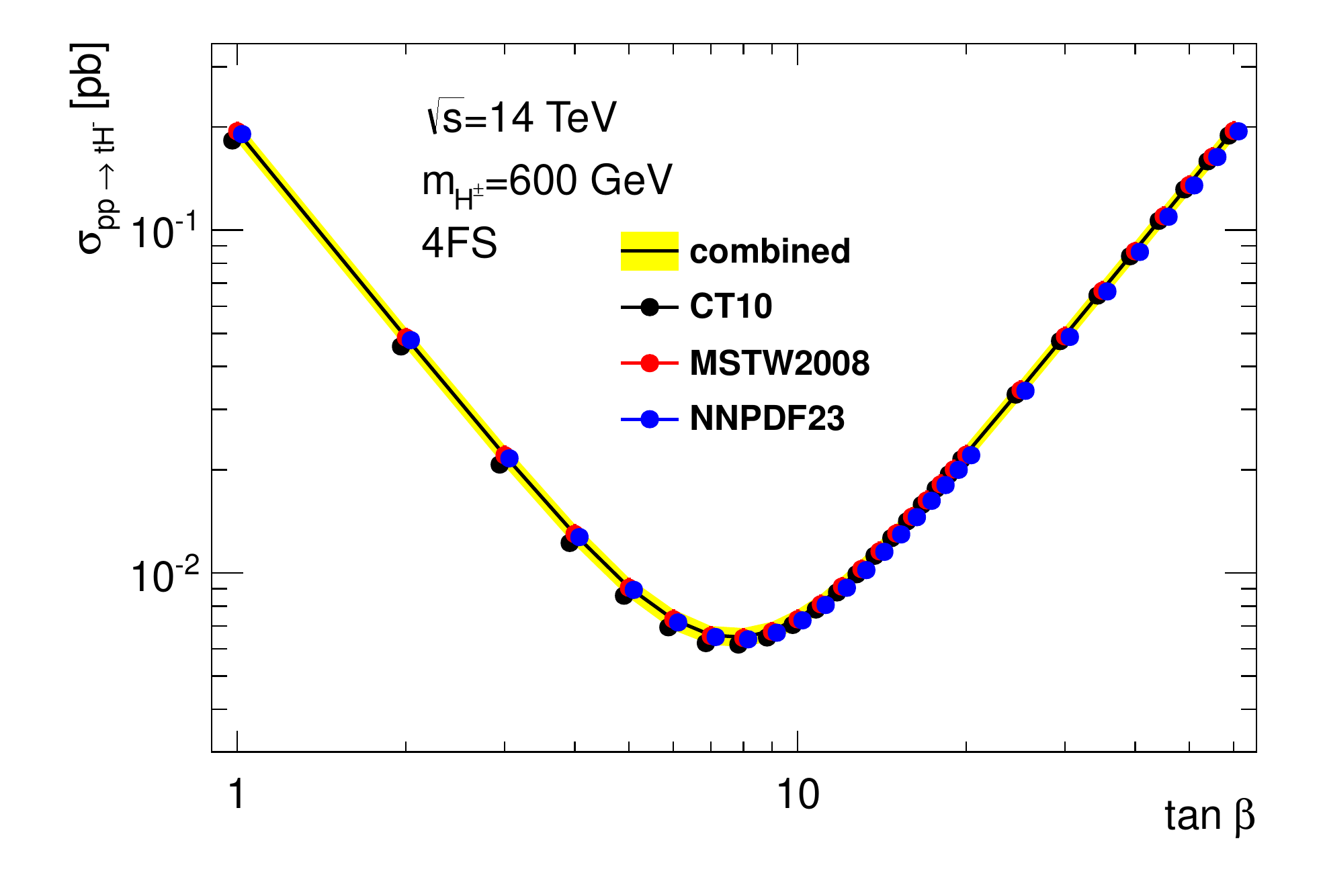}\\
  \includegraphics[width=0.48\textwidth]{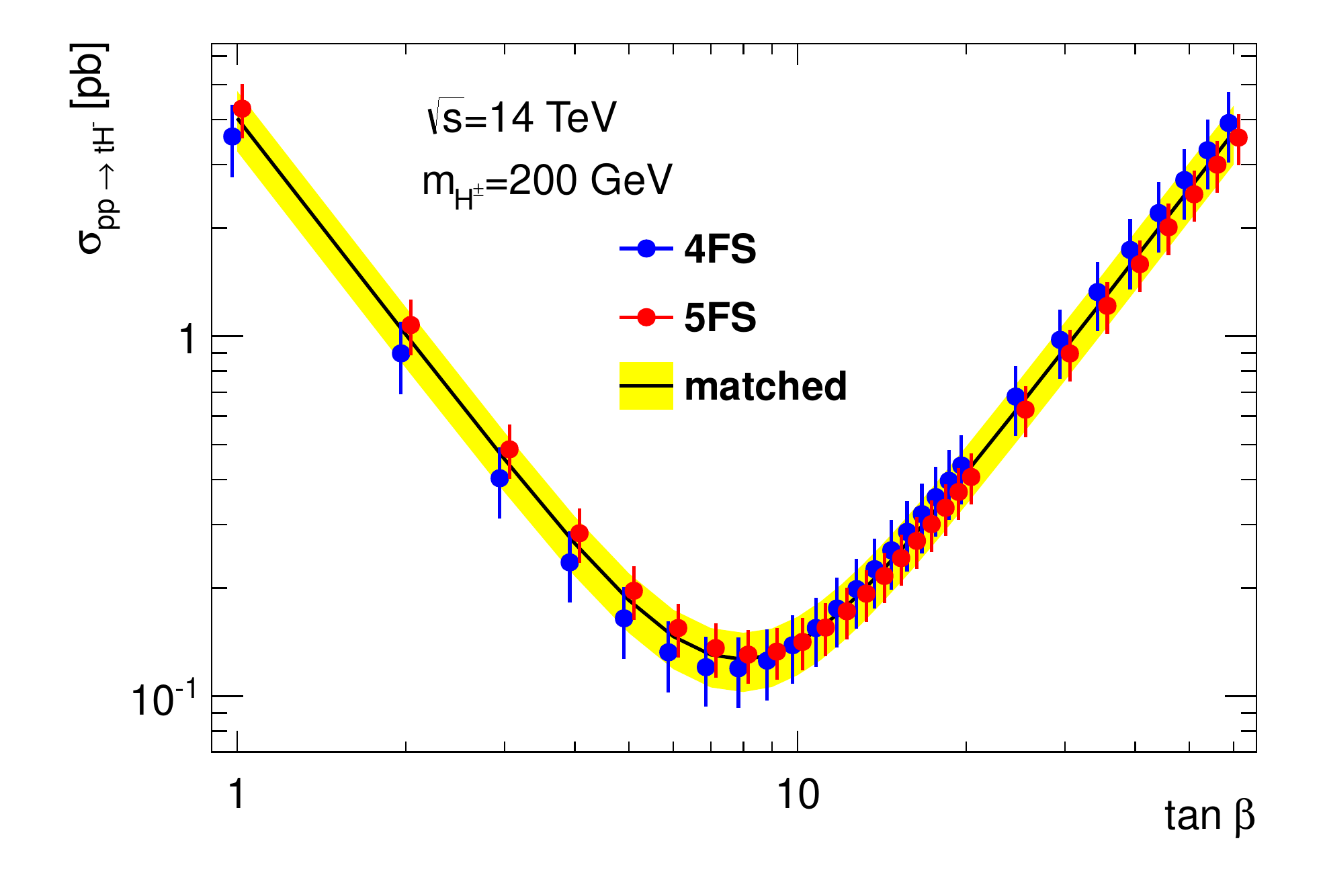}
  \includegraphics[width=0.48\textwidth]{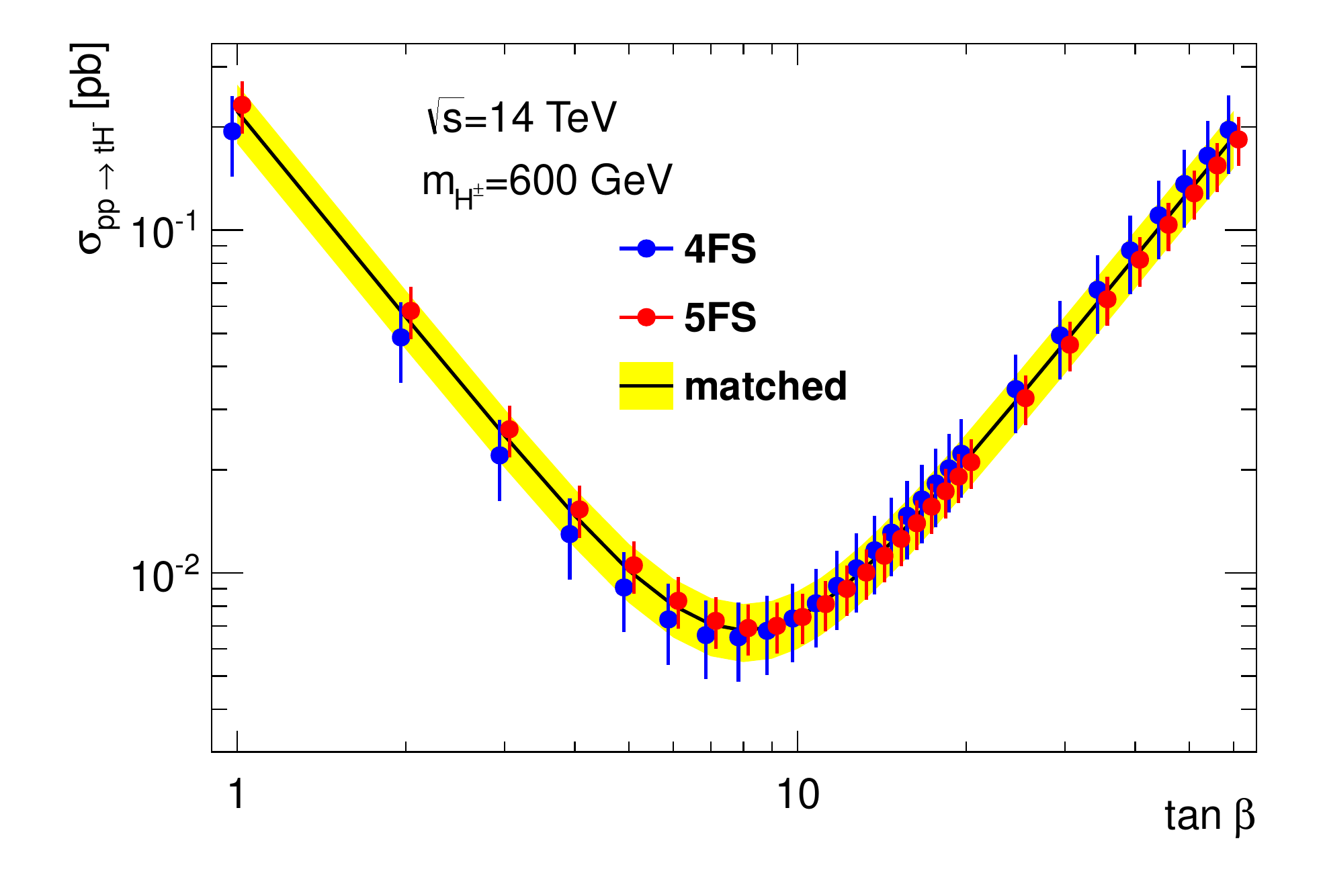}
\end{center}
        \caption{The $pp \rightarrow tH^- + X$ cross section as a function of $\tan \beta$. Shown are the 5FS calculation (top row), 
the 4FS calculation (middle row) and 
the matched calculation (bottom row) for the LHC at $\sqrt{s}=14$ \,TeV for a charged Higgs boson mass of
200\,GeV (left column) and 600\,GeV (right column). The upper two rows 
show the PDF+$\alpha_s$+$m_b$ uncertainties while for the bottom row, the scale uncertainties 
are included as well.}
        \label{fig:scanb}
\end{figure}

The NLO cross sections in the four- and five-flavor schemes and in the Santander-matched 
calculation are displayed in Fig.~\ref{fig:scanb} for the LHC at $\sqrt{s}=14$\,TeV.
The total cross section is essentially proportional to the size of the $tbH^\pm$
coupling which has a minimum for $\tan\beta \approx 8$. 
Comparing the four- andfive-flavor scheme calculations, both agree over  
the whole range of $\tan\beta$ although the difference in the central values 
is slightly larger for small $\tan\beta$. In this region, the results become sensitive 
to the top-bottom-Yukawa interference term $\propto m_t\, m_b$, 
which is absent in the 5FS calculation. 

In a type-I 2HDM all quarks couple to only one of the Higgs doublets. 
In such models, the Yukawa coupling of the charged Higgs boson $H^-$ to a 
top quark and bottom antiquark is given by 
\begin{equation}
g_{t\bar{b}H^-}|_{\rm type-I} = \sqrt{2}\left(\frac{m_t}{v}P_R\cot\beta - \frac{m_b}{v}P_L\cot\beta\right).
\label{eq:tbhcoupling2}
\end{equation}
In contrast to the type-II 2HDM, for type I the bottom Yukawa coupling is not enhanced by $\tan\beta$, 
so that $g_{t\bar{b}H^-}|_{\rm type-I} = \sqrt{2}\, m_t/ v \,P_R\cot\beta + {\cal O}(m_b/m_t)$. 
Up to corrections suppressed by ${\cal O}(m_b/m_t)$, the cross section for heavy charged Higgs boson production in the 
type-I 2HDM, $\sigma |_{\rm type-I} \propto g^2_{t\bar{b}H^-}|_{\rm type-I} \propto 2 (m_t/v)^2 \cot^2\beta + {\cal O}(m_b/m_t)$, 
can thus be obtained 
from the type-II cross section, $\sigma |_{\rm type-II, \tan\beta = 1} \propto g^2_{t\bar{b}H^-}|_{\rm type-II, \tan\beta = 1} \propto 2 (m_t/v)^2 + {\cal O}(m_b/m_t)$, evaluated at $\tan\beta = 1$ and rescaled by $\cot^2\beta$. This relation is correct to all orders in QCD, but \textit{not} to all orders in the electroweak corrections. Given the overall theoretical uncertainty of the cross section prediction of ${\cal O}(30\%)$ it is, however, an excellent approximation and sufficient for all practical purposes. Note that the charged Higgs boson cross section predictions for the type-I and type-II 2HDMs also hold for the so-called lepton-specific and flipped 2HDMs, respectively; see e.g. Ref.~\cite{Branco:2011iw}.

\section{Conclusions}
\label{sec:conclusions}
An updated and improved NLO-QCD calculation of the associated 
production of a heavy charged Higgs boson at the LHC within a type-II two-Higgs-doublet model has been presented. 
The improvements with respect to previous NLO predictions include adopting a new scale setting procedure for the five-flavour scheme, 
a thorough treatment of the theoretical uncertainties based on state-of-the-art PDF sets, 
and a matched prediction for the four- and five-flavour scheme calculations. 

The dynamical choice of the factorization scale in the five-flavour scheme 
calculation significantly improves the agreement between the four- and five-flavour schemes. 
The overall uncertainty 
of the matched cross-section prediction is approximately 20--30\%, and includes the dependence on
 the renormalization scale, the factorization scale, the 
scale of the running bottom quark mass in the Yukawa coupling, as well as the input parameter uncertainties 
in the parton distribution functions, in the QCD coupling $\alpha_s$, and in the bottom quark mass. 
The scale dependence and the input parameter uncertainties contribute about equally to the overall uncertainty. 

The NLO-QCD cross section prediction is provided as a function of $m_{H^\pm}$ and $\tan\beta$, and a 
simple yet accurate prescription is presented to 
convert the result to the production of heavy charged Higgs bosons in type-I , and in 
so-called lepton-specific and flipped two-Higgs-doublet models. The numerical results are made 
available through the wiki page of the {LHC Higgs cross section working group}\cite{HXSwiki}, and allow the interpretation of LHC 
searches for heavy charged Higgs bosons for a wide class of beyond-the-SM scenarios. 

\section*{Acknowledgements}
The authors would like to thank Stefan Dittmaier, Fabio Maltoni, Tilman Plehn and Giovanni Ridolfi for useful discussions and comments. 
MK is grateful to SLAC and
Stanford University for their hospitality. 
The work of MK has been supported by the DFG SFB/TR9 "Computational Particle Physics" and by the U.S.\ Department of Energy under contract DE-AC02-76SF00515. The work of M.S. is supported in part by the European Commission through the HiggsTools Initial Training Network PITN-GA-2012-316704. 
The work of M.U. is supported by the UK Science and Technology Facilities Council.

\bibliography{cH}

\end{document}